\documentclass[letter]{aa}  

\usepackage{graphicx}
\usepackage{txfonts}
\usepackage{inputenc}
\usepackage[]{hyperref}
\hypersetup{colorlinks=true, allcolors=blue}
\usepackage{multicol}
\usepackage{multirow}
\usepackage{siunitx}
\DeclareSIUnit{\solarmass}{\text{M}_{\odot}}
\DeclareSIUnit{\parsec}{pc}
\DeclareSIUnit{\year}{yr}
\DeclareSIUnit{\mag}{mag}
\DeclareSIUnit{\arcsec}{arcsec}
\usepackage{amsmath}
\usepackage{xcolor}
\usepackage{booktabs}

\makeatletter
\renewcommand*\aa@pageof{, page \thepage{} of \pageref*{LastPage}}
\makeatother

\begin{document} 

   \title{The Corona Australis star formation complex is accelerating \\ away from the Galactic plane}

   \author{Laura Posch \inst{1},
          Núria Miret-Roig \inst{1},
          Jo\~ao Alves \inst{1},
          Sebastian Ratzenb{\"o}ck \inst{1,2},
          Josefa Gro{\ss}schedl \inst{1},
          Stefan Meingast \inst{1},
          Catherine Zucker \inst{3},
          Andreas Burkert \inst{4,5}
          }

   \institute{University of Vienna, Department of Astrophysics,
              T\"urkenschanzstrasse 17, 1180 Vienna, Austria
         \and
              University of Vienna, Research Network Data Science at Uni Vienna, Kolingasse 14-16, 1090 Wien, Austria
         \and
              Center for Astrophysics $\mid$ Harvard \& Smithsonian, 60 Garden St., Cambridge, MA, USA 02138
         \and
              University Observatory Munich (USM), Scheinerstrasse 1, D-81679 Munich, Germany
         \and
              Max-Planck-Institut für extraterrestrische Physik (MPE), Giessenbachstr. 1, D-85748 Garching, Germany
        }

   \date{Received xxxx; accepted xxxx}
 
    \abstract
    {
    We study the kinematics of the recently discovered Corona Australis (CrA) chain of clusters by examining the 3D space motion of its young stars using \textit{Gaia} DR3 and APOGEE-2 data. While we observe linear expansion between the clusters in the Cartesian \textit{XY} directions, the expansion along \textit{Z} exhibits a curved pattern. To our knowledge, this is the first time such a nonlinear velocity-position relation has been observed for stellar clusters. We propose a scenario to explain our findings, in which the observed gradient is caused by stellar feedback, accelerating the gas away from the Galactic plane. A traceback analysis confirms that the CrA star formation complex was located near the central clusters of the Scorpius Centaurus (Sco-Cen) OB association 10--15 Myr ago. It contains massive stars and thus offers a natural source of feedback. Based on the velocity of the youngest unbound CrA cluster, we estimate that a median number of about two supernovae would have been sufficient to inject the present-day kinetic energy of the CrA molecular cloud. This number agrees with that of recent studies. The head-tail morphology of the CrA molecular cloud further supports the proposed feedback scenario, in which a feedback force pushed the primordial cloud from the Galactic north, leading to the current separation of 100~pc from the center of Sco-Cen. The formation of spatially and temporally well-defined star formation patterns, such as the CrA chain of clusters, is likely a common process in massive star-forming regions.
    }


   \titlerunning{The Corona Australis star formation complex is accelerating away from the Galactic plane}
   \authorrunning{Posch et al.}

   \maketitle
   

\defcitealias{Ratzenboeck_23a_AA}{R23a}
\defcitealias{Ratzenboeck_23b_AA}{R23b}


\section{Introduction}\label{sec:inroduction}

    Knowing the 3D space motion of molecular clouds is critical for understanding their evolution and the process of star formation.
    Stellar feedback, such as winds from massive stars or supernova (SN) explosions, injects momentum into the surrounding interstellar medium (ISM) and significantly influences the motion of interstellar clouds \citep[e.g.,][]{McCray_83, Walch_15_MNRAS, Girichidis_18_MNRAS}.
    Measuring the 3D velocity of the molecular cloud and estimating its mass allows a direct measurement of the cloud momentum. Assuming that the momentum is predominantly a result of feedback for clouds located near young and massive star-forming regions, this method can help determine the amount of feedback energy deposited onto the cloud. Consequently, it provides a clearer understanding of the role of feedback in the propagation of star formation in molecular clouds \citep[e.g.,][]{Herrington_23_MNRAS}.
    With \textit{Gaia} \citep{Gaia_16_AA} data and complementary data providing additional radial velocity measurements, such as APOGEE\footnote{the Apache Point Observatory Galactic Evolution Experiment} \citep{ApogeeOverview_17_AJ}, we can now trace the motion of molecular clouds by tracing the motion of young stars that are still connected to their parental clouds.
    This method is based on the observation that young stars preserve the motion of their primordial molecular cloud \citep[e.g.,][]{Tobin_09_ApJ, Hacar_16_AA}. This observation was confirmed by \cite{Grossschedl_21_AA} in the Orion southern star formation complex, where they employed this approach to compute cloud orbits using the motion of young clusters. They found coherent cloud motions at the \SI{100}{\parsec} scale, suggesting that feedback plays a significant role in shaping this well-known star-forming region.
    
    The Corona Australis (CrA) star formation complex is one of the molecular clouds that lies closest to Earth, at a distance of about \SI{150}{\parsec} \citep{Zucker_20_AA}. Its densest region is actively forming stars and harbors a deeply embedded cluster, the Coronet cluster \citep{Taylor_Storey_84_MNRAS, Neuhauser_Forbich_08}. This cluster hosts at least one Class\,0 source and several prestellar cores \citep{Sandell_21_ApJ, Bresnahan_18_AA}. \cite{Galli_20_AA} revisited the CrA region using \textit{Gaia} Data Release 2 (DR2) data and identified two kinematically and spatially distinct subgroups of young stars, named ``off-'' and ``on-cloud'' populations. The off-cloud population is located toward the Galactic north of the complex, and the historically known on-cloud population is concentrated close to the densest regions of CrA, in which the Coronet cluster is still embedded. Furthermore, \cite{Esplin_22_AJ} compiled a catalog of young stellar objects (YSOs) in CrA containing off- and on-cloud stars. They found that the clusters are expanding in 3D velocity and position. We investigate this finding in more detail in this letter. 
    In a study of the immediate solar neighborhood, the clustering algorithm developed by \cite{Kerr_21_ApJ} identified CrA without distinguishing between its two subgroups.

    Recently, \cite{Ratzenboeck_23a_AA} (hereafter \citetalias{Ratzenboeck_23a_AA}) published the most complete stellar census of the Scorpio-Centaurus (Sco-Cen) OB association \citep{Blaauw_46_PGro} to date. Sco-Cen is the OB association closest to the Sun. The authors developed a new clustering technique, \texttt{SigMA}, which they applied to \textit{Gaia} DR3 5D astrometric data of stars surrounding Sco-Cen to find more than 13\,000 sources arranged in 37 coeval and comoving clusters\footnote{
    In this letter, we use the word ``cluster'' in the statistical sense, namely, an enhancement over a background, like in \citetalias{Ratzenboeck_23a_AA}. This avoids creating a new term for the spatial/kinematical coherent structures in CrA. None of the CrA clusters, except the embedded Coronet cluster, are expected to be gravitationally bound.}.
    Due to the sensitivity of their algorithm, \citetalias{Ratzenboeck_23a_AA} were able to uncover the substructure in CrA, of which two clusters, named CrA Main and North, share the majority of stellar members with the off- and on-cloud populations found by \cite{Galli_20_AA}, respectively.
    \cite{Ratzenboeck_23b_AA} (hereafter \citetalias{Ratzenboeck_23b_AA}) determined ages for the 37 clusters and found that they are not randomly distributed inside Sco-Cen, but form long coherent chains with well-defined age gradients. One of these chains of clusters is about \SI{100}{\parsec} long, ends on the CrA star formation complex, and was labeled the CrA chain.
    
    This letter aims to investigate the 3D kinematics of this coherent \SI{100}{\parsec} long structure consisting of seven clusters with ages between \SIrange{8}{17}{\mega\year}. This paper is structured as follows:
    In Sect.~\ref{sec:data}, we describe the data selection. In Sect.~\ref{sec:results}, we present our results, and we discuss them in Sect.~\ref{sec:discussion}. Finally, we summarize our findings in Sect.~\ref{sec:conclusion}. Detailed information on our methods can be found in the appendix.

\section{Data}\label{sec:data}

    \begin{figure}[t]
        \centering
        \includegraphics[width=\columnwidth]{Figures/3D_Volumn_Figure.png}
        \caption{Volume plot of the 3D spatial distribution of the CrA chain of clusters as identified by \citetalias{Ratzenboeck_23b_AA}. An online interactive version of the figure is available \protect\href{https://homepage.univie.ac.at/laura.posch/wp-content/uploads/2023/06/ScorpiusCentaurus_Ages_Ratzenboeck_2023.html}{here}. The clusters are color-coded by age. From bottom left to top right and youngest to oldest, they are CrA~Main, CrA~North, Sco~Sting, Sco~Body, V1062~Sco, $\eta$~Lupus, and $\phi$~Lupus.}
        \label{fig:volume}
    \end{figure}

    Our initial sample contains all the stars identified in \citetalias{Ratzenboeck_23a_AA} within the CrA chain of clusters. These clusters in increasing age order are CrA~Main, CrA~North, Scorpio Sting (Sco~Sting), Scorpio Body (Sco~Body), V1062~Sco \citep{Roeser_18_AA}, $\eta$~Lupus, and $\phi$~Lupus.
    In Figure~\ref{fig:volume}, the position and distribution of stellar members for all clusters in our analysis are presented as 3D volumes, each color-coded according to their respective ages.\footnote{Interactive online version of the 3D Figure can be viewed \href{https://homepage.univie.ac.at/laura.posch/wp-content/uploads/2023/06/ScorpiusCentaurus_Ages_Ratzenboeck_2023.html}{here}.} A summary of the spatial and kinematic properties for all seven clusters is available in Table~\ref{tab:properties}.
    The velocity of Sco~Body differs considerably from the rest of the groups in the CrA chain, especially in the azimuthal and vertical directions. We consequently rejected it as a member of the CrA chain. 
    
    We constructed a sample of high-confidence cluster members. We used the stability criterion\footnote{A measure of how frequently individual sources appear across multiple clustering solutions within the cluster ensemble. Stars with high stability values are found to belong to the same cluster more often.} reported in \citetalias{Ratzenboeck_23a_AA} as a first selection criterion. We adopted \texttt{stability} $>$ 11\% for CrA~Main, North, and Sco~Sting, as indicated in \citetalias{Ratzenboeck_23b_AA} for stable cluster members.
    We applied a more restrictive stability cut of \texttt{stability} $>$ 25\% for the other clusters to reduce the number of kinematic and spatial outliers.
    
    To study the kinematics of our cluster selection, we combined the \textit{Gaia} 5D astrometry with radial velocities from \textit{Gaia} DR3 \citep{Gaia_23_AA, Katz_23_AA} and APOGEE-2 from the Sloan Digital Sky Survey (SDSS) DR17 \citep{Apogee2_22_ApJ}. A comprehensive description of the data selection can be found in Appendix~\ref{app:data}. The data selection was designed to obtain a precise 3D velocity sample, which is essential for a detailed kinematic analysis. We transformed the \textit{Gaia} and APOGEE-2 measurements to heliocentric Galactic Cartesian coordinates and velocities (\textit{X,~Y,~Z,~U,~V,~W}). The velocity and position coordinates \textit{U} and \textit{X} correspond to the Galactic radial direction and increase toward the Galactic center, \textit{V} and \textit{Y} correspond to the azimuthal axis and increase along Galactic rotation, and \textit{W} and \textit{Z} correspond to the vertical axis, increasing toward the Galactic north pole.

\section{Results}\label{sec:results}
    
    \begin{figure}[ht!]
        \centering
        \includegraphics[width=\columnwidth]{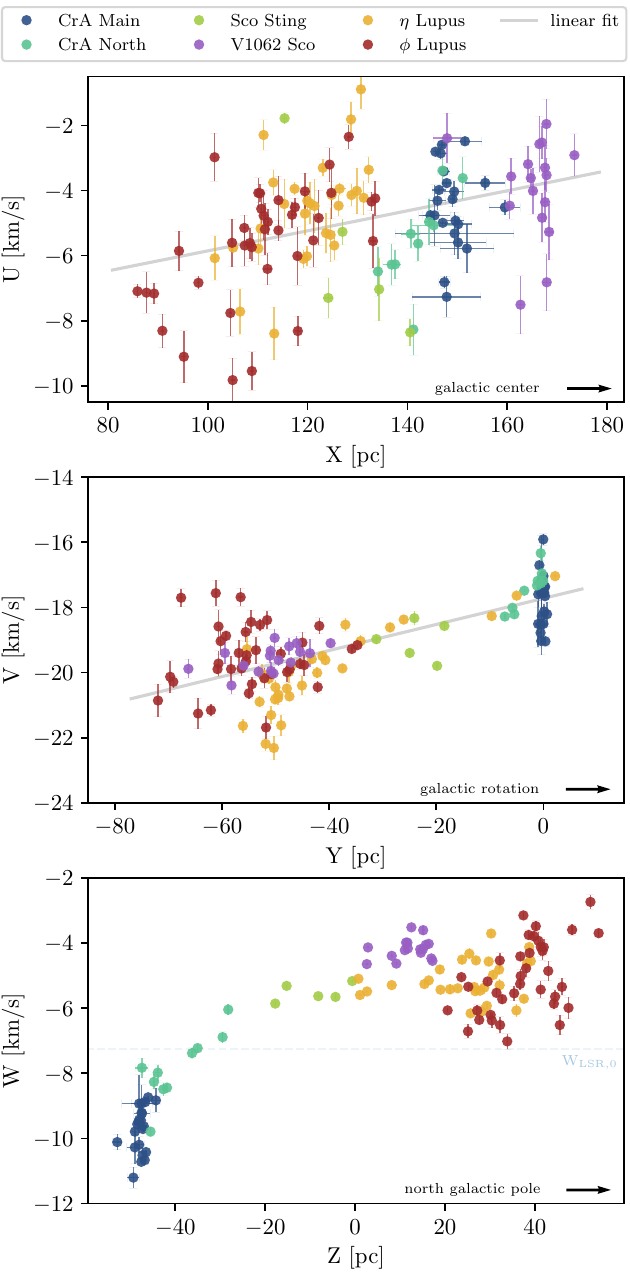}
        \caption{Position-velocity diagrams of the CrA chain of clusters, excluding Sco~Body, in heliocentric Cartesian positions and velocities. We plot regression lines in solid gray and list the fitting parameters in Table~\ref{tab:fit_param}. In the bottom panel, we indicate \textit{W$_{LSR, 0}$} for comparison, and the Galactic plane is at about $Z=-20.8$~pc, based on \cite{Bennett_Bovy_19_MNRAS}.
        }
        \label{fig:gradient}
    \end{figure}

    We computed the cluster orbits of the CrA chain with \texttt{galpy} \citep{galpy_15}; see Appendix~\ref{app:traceback} for details. We find that the younger cluster in the CrA chain, CrA~Main, North, and Sco~Sting, originated close to the core of Sco-Cen. This confirms previous work finding the CrA star formation complex to be part of the Sco-Cen association \citep[e.g.,][]{Mamajek_01_ASPC, Kerr_21_ApJ, Ratzenboeck_23a_AA}.
    
    In Figure~\ref{fig:gradient}, we plot the heliocentric 3D velocity of stars as a function of their heliocentric position. Gravitationally unbound systems born with an initial dispersion in velocities will expand linearly over time as a consequence of their initial velocity dispersion. Galactic differential rotation and disk heating also contribute to the cluster expansion \citep[e.g.,][]{Blaauw_56_ApJ, Asiain_99_AA, Torres_06_AA, Fernandez_08_AA, Mamajek_14_MNRAS, Kuhn_19_ApJ, MiretRoig_20_AA}. To quantify the expansion, we applied a Bayesian linear regression routine to the data (see Appendix~\ref{app:model_comparison} for details), and we report the coefficients in Table~\ref{tab:fit_param}.
    
    We observe linear expansion in both the \textit{U}~versus~\textit{X} and \textit{V}~versus~\textit{Y} spaces. The \textit{U} and \textit{X} uncertainties are larger on average because they are dominated by the radial velocity and parallax measurement uncertainties, respectively (for details, see Appendix~\ref{app:data}). We compared these linear expansion gradients of the CrA chain to gradients derived from the stellar velocity field in the solar vicinity \citep{Nelson_Widrow_22_MNRAS}. We find that the CrA chain is expanding faster than predicted if its motion were dominated solely by Galactic rotation (see Appendix \ref{app:model_comparison} for details).
    
    A visual inspection of Fig.~\ref{fig:gradient} reveals that the \textit{W}~versus ~\textit{Z} relation does not follow a linear expansion (bottom panel of Fig.~\ref{fig:gradient}). The linear fit does not represent the observed data trend well. Instead, the data follow a curved function, implying a complex motion along the vertical axis. The youngest clusters, CrA~Main and North, deviate most from a linear position-velocity gradient, with CrA~Main moving faster to the Galactic south pole than CrA~North.
    We discuss the source of this curved position-velocity trend of the CrA chain in the following section.

    Figure~\ref{fig:planck} shows the morphology of the CrA molecular cloud as observed by Planck \citep{Planck_14_AA}, which exhibits a distinct comet-like structure. The head of the cloud contains the Coronet cluster and CrA Main, and the tail, which extends in the opposite direction of the CrA chain, is largely devoid of star formation activity.   
    
\section{Discussion}\label{sec:discussion}

    We propose that the curved position-velocity trend in the \textit{W}~versus~\textit{Z} space can be explained by an external force that accelerates the motion of the molecular cloud. This scenario is supported by the younger cluster, CrA~Main, which lies farther below the Galactic plane, but moves away from it faster than CrA~North. Without external forces in the Galactic potential, the cloud would slow down after crossing the Galactic plane. This deceleration would result in clusters that are positioned farther from the plane moving away from it at a slower pace, which is in contrast to what we observe.
    Additionally, the shape of the CrA molecular cloud suggests the influence of a force originating from the direction of the Sco-Cen OB association.
    
    Consequently, our interpretation is that we observe a result of stellar feedback, although proper numerical modeling of the motion in the CrA chain is needed to fully understand the origin of the position-velocity pattern. This will be addressed in a follow-up study.

    \subsection{Gradient resulting from a feedback force}\label{sub:feedback}

\begin{figure}[t]
            \centering
            \includegraphics[width=\columnwidth]{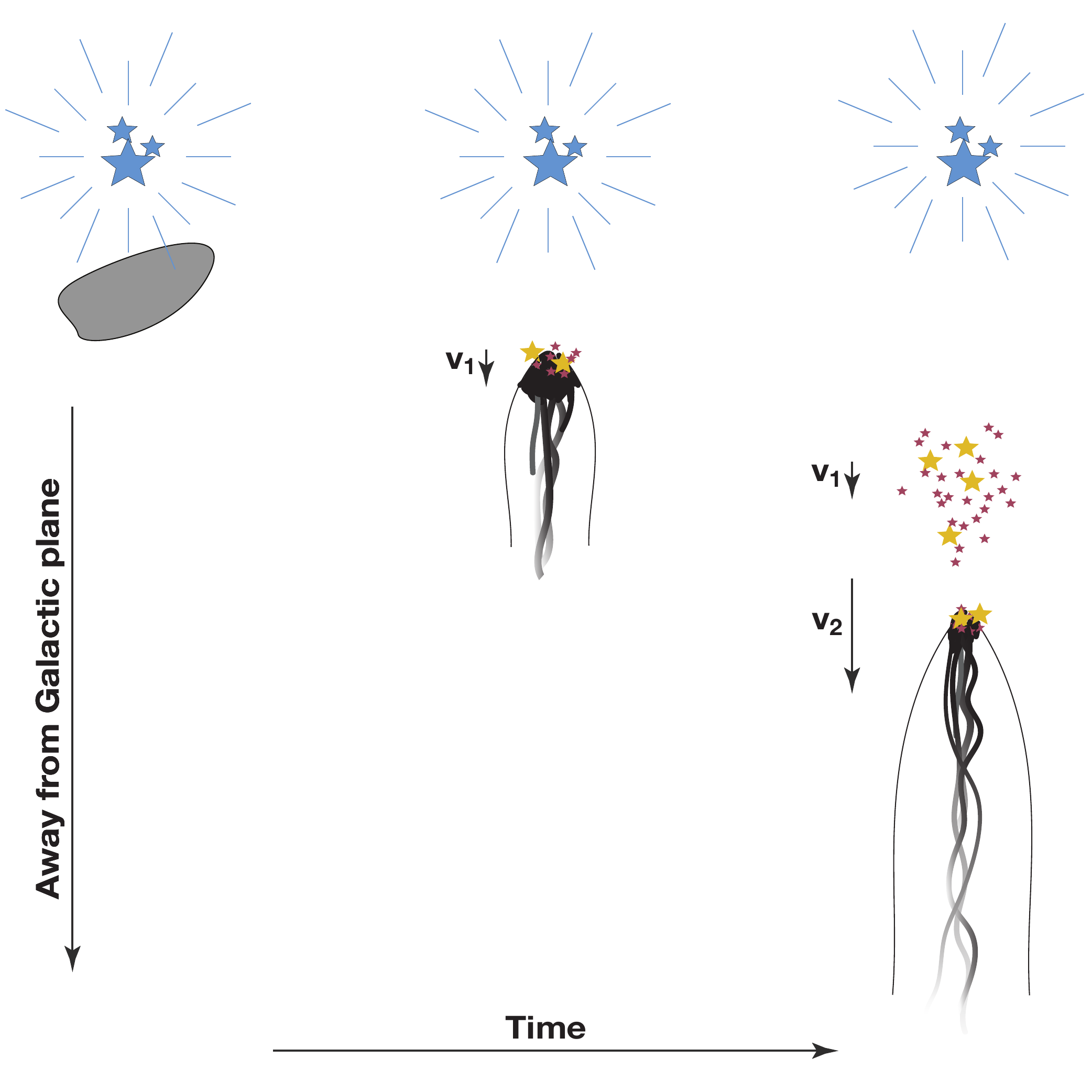}
            \caption{Simplified illustration of the proposed model: A cloud experiences external forces from massive stars, shown in blue, that push it away. This force compresses the cloud, giving it a comet-like shape. Stars then form at the head of the cloud, moving at a velocity v$_1$. As the cloud continues to accelerate, this cycle repeats, leading to the formation of a second group of stars with a higher velocity, v$_2$.}
            \label{fig:sketch}
        \end{figure}

        In the proposed scenario, we consider a cloud that is situated near young clusters containing OB-type stars. In addition to the Galactic gravitational potential, the cloud experiences stellar feedback from massive stars, such as continuous stellar winds and intermittent SN explosions. These forces will accelerate the cloud away from the feedback source. Moreover, the feedback forces will also shape the cloud by compressing the top layers into a dense head and dispersing the gas into a tail. As a result of the compression, the cloud will form stars. Each cluster born from this cloud retains the memory of the cloud motion during cluster formation. After they are born, stars will not feel the feedback forces. At the same time, the remainder of the cloud continues to experience these forces, causing it to accelerate farther away from the feedback source.
        We sketch this scenario in Fig.~\ref{fig:sketch}.
        Multiple clusters born from the same accelerated cloud would create a pattern resembling constant acceleration, forming a quadratic curve in position-velocity space.
        However, because SN feedback occurs intermittently, acceleration is unlikely to be constant.
        In this letter, the term ``cloud acceleration'' is used to describe the increase in momentum resulting from multiple feedback events rather than in the context of the rocket effect, as described in \cite{Oort_Spitzer_55_ApJ}.
        
        In the following paragraphs, we discuss this scenario in more detail and propose that V1062~Sco, $\eta$~Lupus, or $\phi$~Lupus, or a combination of all three, are the likely source of feedback for the formation of the CrA~Main, North, and possibly also of the Sco~Sting clusters.

    \subsection{CrA as part of the Sco-Cen association}\label{sub:environment}

        \begin{figure*}[t]
            \centering
            \includegraphics[width=\textwidth]{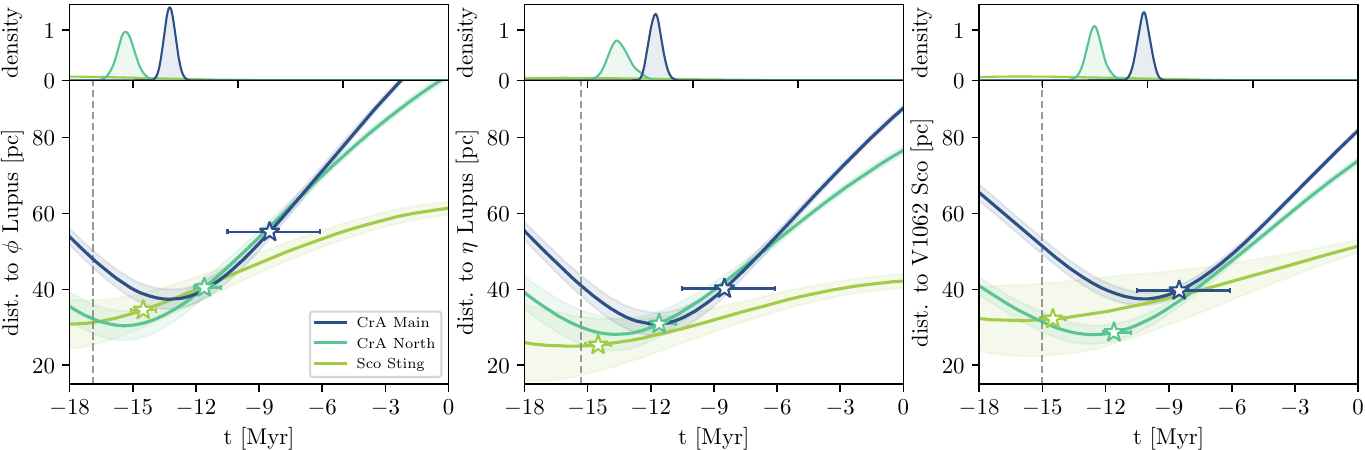}
            \caption{Distance between the older and more massive clusters $\phi$~Lupus, $\eta$~Lupus, and V1062~Sco and each of the younger clusters CrA~Main, North, and Sco~Sting over the past \SI{18}{\mega\year}. Ages and their corresponding uncertainties for CrA~Main, North, and Sco~Sting are indicated with a star. Prior to star formation, the motion of these clusters corresponds to that of their primordial clouds. The ages of the reference clusters are indicated with the vertical dashed lines. All ages are taken from \citetalias{Ratzenboeck_23b_AA} and we list them in Table~\ref{tab:properties}. At the top, we show the distribution of the times of minimum separation; the distribution of Sco~Sting is widespread and hardly visible.}
            \label{fig:traceback}
        \end{figure*}

        The Sco-Cen OB association likely injects feedback into its larger environment, in which the CrA molecular cloud is located.
        In \citetalias{Ratzenboeck_23b_AA}, the authors identified multiple star formation episodes in Sco-Cen, with a main burst of star formation about \SI{15}{\mega\year} ago. Over the past \SI{15}{\mega\year}, around \num{15} SN events occurred in this region, originating from the Sco-Cen OB population \citep{Zucker_22_Natur}. The Sco-Cen OB association currently still harbors more than 100 B-type stars. A recent study lists \num{181} B-type stars as members of the Sco-Cen OB association \citep{Gratton_23_arXiv}. Their membership selection was based on previous studies \citep{Rizzuto_11_MNRAS, Janson_21_AA}. 

        To investigate the relation between the younger and older clusters in the CrA chain, we computed the distances between the cluster centers of CrA~Main, North, and Sco~Sting, and each of the clusters V1062~Sco, $\eta$~Lupus, and $\phi$~Lupus as a function of traceback time. We plot the results in Fig.~\ref{fig:traceback} and list the minimum distances and times of minimum distance in Table~\ref{tab:distances}. We find that the primordial cloud of the three younger clusters converges with the older clusters \SIrange{10}{15}{\mega\year} ago at an average minimum separation of about \SI{30}{\parsec}. Furthermore, we find that the times of minimum distances between the younger clusters and the clusters they encounter correlate with the age of the encountered clusters: The oldest cluster, $\phi$~Lupus, is at its minimum distance earlier than $\eta$~Lupus or V1062~Sco, suggesting a gradual expansion. This can also be seen in the top panels of Fig.~\ref{fig:traceback}, where the distribution of the times of minimum distances peaks sequentially. We note that this result is more uncertain for Sco~Sting because the orbit of this cluster was calculated with only seven cluster members, leading to larger uncertainties. Finally, we find that the onset of star formation in the young group of clusters matches the determined minimum distance to V1062~Sco within the age uncertainties. This alignment is evident in the peak positions of the probability distribution in the upper panels, coinciding with the cluster ages, as denoted by the star markers in the main panels. In contrast, the closest encounters with $\eta$ and $\phi$~Lupus appear to have occurred prior to the onset of star formation. This would allow time for the molecular cloud to contract as a result of feedback acting on the gas before star formation is initiated. 

        We particularly note the most significant impact from feedback along the vertical axis, which indicates that the feedback event primarily operated in this direction. However, we observe the impact in all dimensions because the steeper linear gradients along the \textit{X}- and \textit{Y}-axes compared to the solar neighborhood average can be explained by stellar feedback as well. In a study of the Cygnus OB associations, \cite{Quintana_21_MNRAS} similarly proposed that stellar feedback generated the velocity-position gradient, whereas \cite{Quintana_22_MNRAS} argued that this expansion can also be explained by the turbulence of the primordial cloud. \cite{Drew_21_MNRAS} attributed similarly inclined position-velocity gradients on larger scales to Galactic rotation.
        
        In an attempt to quantify the curved position-velocity pattern in the \textit{W}~versus~\textit{Z} space, we find that a subsection of the CrA chain, namely CrA~Main, CrA~North, and Sco~Sting, can be fit well with a quadratic equation (see Fig.~\ref{fig:quadratic} in Appendix~\ref{app:model_comparison}).
        This quadratic fit includes the younger clusters that were likely born from the cloud during its acceleration.
        However, the feedback force was likely not constant, but varied over time, creating a more complex pattern in the position-velocity space, which is what we find for the CrA chain. Based on this fit, it remains unclear which of the older and more massive clusters, likely having contained multiple massive stars, acted as the primary feedback source. Quantifying the overall trend will require modeling the dynamics of all CrA chain clusters.

        The wind-blown shape of the CrA molecular cloud (see Fig.~\ref{fig:planck}) further indicates influence from an external force that pushes from the Galactic north against the top of the cloud and compresses it, which is the likely reason for the comet-like morphology of the cloud.
        In a study investigating the impact of stellar feedback on dust clumps, \cite{Kirchschlager_23_MNRAS} found shapes very similar to the CrA molecular cloud for large magnetic field strengths and high gas densities. The simulated shock impacts the dust clump by compressing the exposed side, stripping off material from outer shells, blowing it away, accelerating and fragmenting it to form a nonsymmetric structure. Morphological structures similar to the CrA molecular cloud and the dust clumps in \cite{Kirchschlager_23_MNRAS} were found by \cite{Rogers_Pittard_14_MNRAS}. They simulated massive stellar feedback that affects the remnant molecular material following star formation. While these simulations produce structures that are several orders of magnitude smaller than we observe in the CrA star formation complex, we tentatively propose the outcome to scale self-similarly under similar conditions. Further modeling is needed to confirm our assumptions. In both studies, the compressed heads of these comet-like structures point toward the source of feedback. In the case of CrA, the dense head of the molecular cloud points toward the Sco-Cen OB association.

        A study by \cite{Herrington_23_MNRAS} highlights the importance of stellar feedback from previous generations of stars in driving gas motion on larger scales and structuring the ISM. 3D dust maps \citep[e.g.,][]{Leike_20_AA, Edenhofer_23_arXiv} show low-density bubbles ubiquitously distributed in the ISM.
        The shape of the CrA molecular cloud might additionally be affected by compression from the expanding shell described in \cite{Bracco_20_AA}. However, we find that the clusters that could be affected by it, CrA Main and North, do not display obvious kinematic effects caused by this bubble.

    \subsection{Momentum estimation}\label{ssub:momentum}
    
        As described in \cite{Grossschedl_21_AA}, we can estimate the cloud motion from the 3D velocities of its associated young stars and calculate the cloud momentum with its mass. Using the momentum of the CrA molecular cloud, we can make an educated guess on the feedback force needed to accelerate the cloud under the feedback scenario. For a detailed analysis, see Appendix~\ref{app:LoSMom}, where we also evaluate the possibility of momentum from winds of B-type supergiants (Sect.~\ref{ssub:btype}).

        We applied Monte Carlo-type sampling to all parameters and obtained a probability distribution of the necessary number of SNe (see Fig.~\ref{fig:supernovae}).
        First, we calculated the current momentum of the CrA molecular cloud based on the cloud mass and velocity, resulting in a medium momentum of \SI{25000}{\solarmass\km\per\s}. The estimated number of SNe then depends on the distance between the cloud and the SN explosion, and furthermore, on the surface area fraction of the SN sphere hitting the cloud.
        By adopting a total SN momentum of \SIrange{2e5}{4e5}{\solarmass\km\per\s} \citep[e.g.,][]{Iffrig_Hennebelle_15_AA, Kim_Ostriker_15_ApJ, Walch_Naab_15_MNRAS, Haid_16_MNRAS} that is distributed over the surface of the SN sphere, we received a median number of about two SN explosions that are needed to inject the momentum that the CrA molecular cloud has today.
        This rough estimate neglects factors such as additional influence from stellar winds and photoionizing radiation that affects the surrounding ISM and the cloud itself prior to SN explosions.
    
        Based on the findings of \cite{Zucker_22_Natur} and \citetalias{Ratzenboeck_23b_AA}, two SN explosions in close proximity to CrA are accounted for. These observations further support the theory that feedback is a plausible mechanism for accelerating the cloud and producing the observed curved gradient.
            
\section{Conclusions}\label{sec:conclusion}

    We investigate the 3D kinematics of a spatially coherent structure in Sco-Cen, including the clusters CrA~Main, CrA~North, Sco~Sting, $\eta$~Lupus, $\phi$~Lupus, and V1062~Sco. We find a curved position-velocity gradient for the CrA chain of clusters along the vertical axis, \textit{W}~versus~\textit{Z}, instead of linear cluster expansion. To explain this unexpected trend, we propose the following possible scenario.    

    The observables can be reproduced by a scenario in which a molecular cloud is subject to a force pushing from the Galactic north and compressing it. The likely origin of this force is feedback from massive stars, such as SN explosions and stellar winds originating in Sco-Cen. The traceback analysis shows that about \SIrange{10}{15}{\mega\year} ago, CrA~Main, North, and Sco~Sting were at a distance of about \SI{30}{\parsec} to the massive older clusters in the CrA chain, namely V1062~Sco, $\eta$ and $\phi$~Lupus, which are candidates for the origin of stellar feedback. Based on the momentum analysis, our estimation indicates that about two SNe would suffice to account for the current momentum of the CrA molecular cloud and push it approximately \SI{100}{\parsec} beyond the association as a consequence. The wind-blown appearance of the CrA molecular cloud further supports this scenario.

    It is highly likely that chains of young star clusters experiencing acceleration are prevalent in star-forming regions throughout the Milky Way: The Sco-Cen region is a common star-forming OB association, harboring massive stars much like numerous other star-forming regions.
    The motion of the CrA star formation complex, as determined from \textit{Gaia} and APOGEE-2 data, is a powerful probe of the effects of stellar feedback on gas structures. Future measurements in other star formation environments will help constrain the role of feedback in driving star formation. Complete numerical modeling of the motion of the CrA cloud, fed by the observational constraints in this work, will offer insight into the physical conditions that allow the formation of chains of clusters. 
    
\begin{acknowledgements}
We thank an anonymous referee for their helpful comments. 
Co-funded by the European Union (ERC, ISM-FLOW, 101055318). Views and opinions expressed are, however, those of the author(s) only and do not necessarily reflect those of the European Union or the European Research Council. Neither the European Union nor the granting authority can be held responsible for them. 
S.~Ratzenb{\"o}ck acknowledges funding by the Austrian Research Promotion Agency (FFG, \url{https://www.ffg.at/}) under project number FO999892674.
J.\,Gro{\ss}schedl acknowledges funding by the Austrian Research Promotion Agency (FFG) under project number 873708.
This research has made use of \textit{Python}, \url{https://www.python.org}, of Astropy, a community-developed core \textit{Python} package for Astronomy \citep{astropy_18, Astropy_22_ApJ}, NumPy \citep{numpy_20}, Matplotlib \citep{matplotlib_07}, Galpy \citep{galpy_15}, Scipy \citep{Scipy_20_NMeth}, PyMC \citep{PyMC_16}, and Plotly \citep{plotly}.
\end{acknowledgements}
\bibliographystyle{aa.bst}
\bibliography{AA_2023_47186.bib}

\begin{thebibliography}{73}
\expandafter\ifx\csname natexlab\endcsname\relax\def\natexlab#1{#1}\fi

\bibitem[{{Abdurro'uf} {et~al.}(2022){Abdurro'uf}, {Accetta}, {Aerts}, {Silva
  Aguirre}, {Ahumada}, {Ajgaonkar}, {Filiz Ak}, {Alam}, {Allende Prieto},
  {Almeida}, {Anders}, {Anderson}, {Andrews}, {Anguiano}, {Aquino-Ort{\'\i}z},
  {Arag{\'o}n-Salamanca}, {Argudo-Fern{\'a}ndez}, {Ata}, {Aubert},
  {Avila-Reese}, {Badenes}, {Barb{\'a}}, {Barger}, {Barrera-Ballesteros},
  {Beaton}, {Beers}, {Belfiore}, {Bender}, {Bernardi}, {Bershady}, {Beutler},
  {Bidin}, {Bird}, {Bizyaev}, {Blanc}, {Blanton}, {Boardman}, {Bolton},
  {Boquien}, {Borissova}, {Bovy}, {Brandt}, {Brown}, {Brownstein}, {Brusa},
  {Buchner}, {Bundy}, {Burchett}, {Bureau}, {Burgasser}, {Cabang}, {Campbell},
  {Cappellari}, {Carlberg}, {Wanderley}, {Carrera}, {Cash}, {Chen}, {Chen},
  {Cherinka}, {Chiappini}, {Choi}, {Chojnowski}, {Chung}, {Clerc}, {Cohen},
  {Comerford}, {Comparat}, {da Costa}, {Covey}, {Crane}, {Cruz-Gonzalez},
  {Culhane}, {Cunha}, {Dai}, {Damke}, {Darling}, {Davidson}, {Davies},
  {Dawson}, {De Lee}, {Diamond-Stanic}, {Cano-D{\'\i}az}, {S{\'a}nchez},
  {Donor}, {Duckworth}, {Dwelly}, {Eisenstein}, {Elsworth}, {Emsellem},
  {Eracleous}, {Escoffier}, {Fan}, {Farr}, {Feng}, {Fern{\'a}ndez-Trincado},
  {Feuillet}, {Filipp}, {Fillingham}, {Frinchaboy}, {Fromenteau}, {Galbany},
  {Garc{\'\i}a}, {Garc{\'\i}a-Hern{\'a}ndez}, {Ge}, {Geisler}, {Gelfand},
  {G{\'e}ron}, {Gibson}, {Goddy}, {Godoy-Rivera}, {Grabowski}, {Green},
  {Greener}, {Grier}, {Griffith}, {Guo}, {Guy}, {Hadjara}, {Harding},
  {Hasselquist}, {Hayes}, {Hearty}, {Hern{\'a}ndez}, {Hill}, {Hogg},
  {Holtzman}, {Horta}, {Hsieh}, {Hsu}, {Hsu}, {Huber}, {Huertas-Company},
  {Hutchinson}, {Hwang}, {Ibarra-Medel}, {Chitham}, {Ilha}, {Imig}, {Jaekle},
  {Jayasinghe}, {Ji}, {Johnson}, {Jones}, {J{\"o}nsson}, {Katkov}, {Khalatyan},
  {Kinemuchi}, {Kisku}, {Knapen}, {Kneib}, {Kollmeier}, {Kong}, {Kounkel},
  {Kreckel}, {Krishnarao}, {Lacerna}, {Lane}, {Langgin}, {Lavender}, {Law},
  {Lazarz}, {Leung}, {Leung}, {Lewis}, {Li}, {Li}, {Lian}, {Liang}, {Lin},
  {Lin}, {Lin}, {Lintott}, {Long}, {Longa-Pe{\~n}a}, {L{\'o}pez-Cob{\'a}},
  {Lu}, {Lundgren}, {Luo}, {Mackereth}, {de la Macorra}, {Mahadevan},
  {Majewski}, {Manchado}, {Mandeville}, {Maraston}, {Margalef-Bentabol},
  {Masseron}, {Masters}, {Mathur}, {McDermid}, {Mckay}, {Merloni},
  {Merrifield}, {Meszaros}, {Miglio}, {Di Mille}, {Minniti}, {Minsley},
  {Monachesi}, {Moon}, {Mosser}, {Mulchaey}, {Muna}, {Mu{\~n}oz}, {Myers},
  {Myers}, {Nadathur}, {Nair}, {Nandra}, {Neumann}, {Newman}, {Nidever},
  {Nikakhtar}, {Nitschelm}, {O'Connell}, {Garma-Oehmichen}, {Luan Souza de
  Oliveira}, {Olney}, {Oravetz}, {Ortigoza-Urdaneta}, {Osorio}, {Otter},
  {Pace}, {Padilla}, {Pan}, {Pan}, {Parikh}, {Parker}, {Peirani}, {Pe{\~n}a
  Ram{\'\i}rez}, {Penny}, {Percival}, {Perez-Fournon}, {Pinsonneault},
  {Poidevin}, {Poovelil}, {Price-Whelan}, {B{\'a}rbara de Andrade Queiroz},
  {Raddick}, {Ray}, {Rembold}, {Riddle}, {Riffel}, {Riffel}, {Rix}, {Robin},
  {Rodr{\'\i}guez-Puebla}, {Roman-Lopes}, {Rom{\'a}n-Z{\'u}{\~n}iga}, {Rose},
  {Ross}, {Rossi}, {Rubin}, {Salvato}, {S{\'a}nchez}, {S{\'a}nchez-Gallego},
  {Sanderson}, {Santana Rojas}, {Sarceno}, {Sarmiento}, {Sayres}, {Sazonova},
  {Schaefer}, {Schiavon}, {Schlegel}, {Schneider}, {Schultheis}, {Schwope},
  {Serenelli}, {Serna}, {Shao}, {Shapiro}, {Sharma}, {Shen}, {Shetrone}, {Shu},
  {Simon}, {Skrutskie}, {Smethurst}, {Smith}, {Sobeck}, {Spoo}, {Sprague},
  {Stark}, {Stassun}, {Steinmetz}, {Stello}, {Stone-Martinez},
  {Storchi-Bergmann}, {Stringfellow}, {Stutz}, {Su}, {Taghizadeh-Popp},
  {Talbot}, {Tayar}, {Telles}, {Teske}, {Thakar}, {Theissen}, {Tkachenko},
  {Thomas}, {Tojeiro}, {Hernandez Toledo}, {Troup}, {Trump}, {Trussler},
  {Turner}, {Tuttle}, {Unda-Sanzana}, {V{\'a}zquez-Mata}, {Valentini},
  {Valenzuela}, {Vargas-Gonz{\'a}lez}, {Vargas-Maga{\~n}a}, {Alfaro},
  {Villanova}, {Vincenzo}, {Wake}, {Warfield}, {Washington}, {Weaver},
  {Weijmans}, {Weinberg}, {Weiss}, {Westfall}, {Wild}, {Wilde}, {Wilson},
  {Wilson}, {Wilson}, {Wolf}, {Wood-Vasey}, {Yan}, {Zamora}, {Zasowski},
  {Zhang}, {Zhao}, {Zheng}, {Zheng}, \& {Zhu}}]{Apogee2_22_ApJ}
{Abdurro'uf}, {Accetta}, K., {Aerts}, C., {et~al.} 2022, \apjs, 259, 35

\bibitem[{{Alves} {et~al.}(1999){Alves}, {Lada}, \& {Lada}}]{Alves_99_ApJ}
{Alves}, J., {Lada}, C.~J., \& {Lada}, E.~A. 1999, \apj, 515, 265

\bibitem[{{Alves} {et~al.}(2014){Alves}, {Lombardi}, \& {Lada}}]{Alves_14_AA}
{Alves}, J., {Lombardi}, M., \& {Lada}, C.~J. 2014, \aap, 565, A18

\bibitem[{{Asiain} {et~al.}(1999){Asiain}, {Figueras}, \&
  {Torra}}]{Asiain_99_AA}
{Asiain}, R., {Figueras}, F., \& {Torra}, J. 1999, \aap, 350, 434

\bibitem[{{Astropy Collaboration} {et~al.}(2022){Astropy Collaboration},
  {Price-Whelan}, {Lim}, {Earl}, {Starkman}, {Bradley}, {Shupe}, {Patil},
  {Corrales}, {Brasseur}, {N{\"o}the}, {Donath}, {Tollerud}, {Morris},
  {Ginsburg}, {Vaher}, {Weaver}, {Tocknell}, {Jamieson}, {van Kerkwijk},
  {Robitaille}, {Merry}, {Bachetti}, {G{\"u}nther}, {Aldcroft},
  {Alvarado-Montes}, {Archibald}, {B{\'o}di}, {Bapat}, {Barentsen},
  {Baz{\'a}n}, {Biswas}, {Boquien}, {Burke}, {Cara}, {Cara}, {Conroy},
  {Conseil}, {Craig}, {Cross}, {Cruz}, {D'Eugenio}, {Dencheva}, {Devillepoix},
  {Dietrich}, {Eigenbrot}, {Erben}, {Ferreira}, {Foreman-Mackey}, {Fox},
  {Freij}, {Garg}, {Geda}, {Glattly}, {Gondhalekar}, {Gordon}, {Grant},
  {Greenfield}, {Groener}, {Guest}, {Gurovich}, {Handberg}, {Hart},
  {Hatfield-Dodds}, {Homeier}, {Hosseinzadeh}, {Jenness}, {Jones}, {Joseph},
  {Kalmbach}, {Karamehmetoglu}, {Ka{\l}uszy{\'n}ski}, {Kelley}, {Kern},
  {Kerzendorf}, {Koch}, {Kulumani}, {Lee}, {Ly}, {Ma}, {MacBride}, {Maljaars},
  {Muna}, {Murphy}, {Norman}, {O'Steen}, {Oman}, {Pacifici}, {Pascual},
  {Pascual-Granado}, {Patil}, {Perren}, {Pickering}, {Rastogi}, {Roulston},
  {Ryan}, {Rykoff}, {Sabater}, {Sakurikar}, {Salgado}, {Sanghi}, {Saunders},
  {Savchenko}, {Schwardt}, {Seifert-Eckert}, {Shih}, {Jain}, {Shukla}, {Sick},
  {Simpson}, {Singanamalla}, {Singer}, {Singhal}, {Sinha}, {Sip{\H{o}}cz},
  {Spitler}, {Stansby}, {Streicher}, {{\v{S}}umak}, {Swinbank}, {Taranu},
  {Tewary}, {Tremblay}, {Val-Borro}, {Van Kooten}, {Vasovi{\'c}}, {Verma}, {de
  Miranda Cardoso}, {Williams}, {Wilson}, {Winkel}, {Wood-Vasey}, {Xue},
  {Yoachim}, {Zhang}, {Zonca}, \& {Astropy Project
  Contributors}}]{Astropy_22_ApJ}
{Astropy Collaboration}, {Price-Whelan}, A.~M., {Lim}, P.~L., {et~al.} 2022,
  \apj, 935, 167

\bibitem[{{Astropy Collaboration} {et~al.}(2018){Astropy Collaboration},
  {Price-Whelan}, {Sip{\H{o}}cz}, {G{\"u}nther}, {Lim}, {Crawford}, {Conseil},
  {Shupe}, {Craig}, {Dencheva}, {Ginsburg}, {Vand erPlas}, {Bradley},
  {P{\'e}rez-Su{\'a}rez}, {de Val-Borro}, {Aldcroft}, {Cruz}, {Robitaille},
  {Tollerud}, {Ardelean}, {Babej}, {Bach}, {Bachetti}, {Bakanov}, {Bamford},
  {Barentsen}, {Barmby}, {Baumbach}, {Berry}, {Biscani}, {Boquien}, {Bostroem},
  {Bouma}, {Brammer}, {Bray}, {Breytenbach}, {Buddelmeijer}, {Burke},
  {Calderone}, {Cano Rodr{\'\i}guez}, {Cara}, {Cardoso}, {Cheedella}, {Copin},
  {Corrales}, {Crichton}, {D'Avella}, {Deil}, {Depagne}, {Dietrich}, {Donath},
  {Droettboom}, {Earl}, {Erben}, {Fabbro}, {Ferreira}, {Finethy}, {Fox},
  {Garrison}, {Gibbons}, {Goldstein}, {Gommers}, {Greco}, {Greenfield},
  {Groener}, {Grollier}, {Hagen}, {Hirst}, {Homeier}, {Horton}, {Hosseinzadeh},
  {Hu}, {Hunkeler}, {Ivezi{\'c}}, {Jain}, {Jenness}, {Kanarek}, {Kendrew},
  {Kern}, {Kerzendorf}, {Khvalko}, {King}, {Kirkby}, {Kulkarni}, {Kumar},
  {Lee}, {Lenz}, {Littlefair}, {Ma}, {Macleod}, {Mastropietro}, {McCully},
  {Montagnac}, {Morris}, {Mueller}, {Mumford}, {Muna}, {Murphy}, {Nelson},
  {Nguyen}, {Ninan}, {N{\"o}the}, {Ogaz}, {Oh}, {Parejko}, {Parley}, {Pascual},
  {Patil}, {Patil}, {Plunkett}, {Prochaska}, {Rastogi}, {Reddy Janga},
  {Sabater}, {Sakurikar}, {Seifert}, {Sherbert}, {Sherwood-Taylor}, {Shih},
  {Sick}, {Silbiger}, {Singanamalla}, {Singer}, {Sladen}, {Sooley},
  {Sornarajah}, {Streicher}, {Teuben}, {Thomas}, {Tremblay}, {Turner},
  {Terr{\'o}n}, {van Kerkwijk}, {de la Vega}, {Watkins}, {Weaver}, {Whitmore},
  {Woillez}, {Zabalza}, \& {Astropy Contributors}}]{astropy_18}
{Astropy Collaboration}, {Price-Whelan}, A.~M., {Sip{\H{o}}cz}, B.~M., {et~al.}
  2018, \aj, 156, 123

\bibitem[{{Bennett} \& {Bovy}(2019)}]{Bennett_Bovy_19_MNRAS}
{Bennett}, M. \& {Bovy}, J. 2019, \mnras, 482, 1417

\bibitem[{{Blaauw}(1946)}]{Blaauw_46_PGro}
{Blaauw}, A. 1946, Publications of the Kapteyn Astronomical Laboratory
  Groningen, 52, 1

\bibitem[{{Blaauw}(1956)}]{Blaauw_56_ApJ}
{Blaauw}, A. 1956, \apj, 123, 408

\bibitem[{Bovy(2015)}]{galpy_15}
Bovy, J. 2015, The Astrophysical Journal Supplement Series, 216, 29

\bibitem[{Bracco {et~al.}(2020)Bracco, Bresnahan, Palmeirim, Arzoumanian,
  André, Ward-Thompson, \& Marchal}]{Bracco_20_AA}
Bracco, A., Bresnahan, D., Palmeirim, P., {et~al.} 2020, Astronomy \&
  Astrophysics, 644, A5

\bibitem[{{Bresnahan} {et~al.}(2018){Bresnahan}, {Ward-Thompson}, {Kirk},
  {Pattle}, {Eyres}, {White}, {K{\"o}nyves}, {Men'shchikov}, {Andr{\'e}},
  {Schneider}, {Di Francesco}, {Arzoumanian}, {Benedettini}, {Ladjelate},
  {Palmeirim}, {Bracco}, {Molinari}, {Pezzuto}, \&
  {Spinoglio}}]{Bresnahan_18_AA}
{Bresnahan}, D., {Ward-Thompson}, D., {Kirk}, J.~M., {et~al.} 2018, \aap, 615,
  A125

\bibitem[{{Casey} {et~al.}(1998){Casey}, {Mathieu}, {Vaz}, {Andersen}, \&
  {Suntzeff}}]{Casey_98_AJ}
{Casey}, B.~W., {Mathieu}, R.~D., {Vaz}, L. P.~R., {Andersen}, J., \&
  {Suntzeff}, N.~B. 1998, \aj, 115, 1617

\bibitem[{{Dame} {et~al.}(2001){Dame}, {Hartmann}, \& {Thaddeus}}]{Dame_01_ApJ}
{Dame}, T.~M., {Hartmann}, D., \& {Thaddeus}, P. 2001, \apj, 547, 792

\bibitem[{{de Geus} {et~al.}(1989){de Geus}, {de Zeeuw}, \&
  {Lub}}]{deGeus_89_AA}
{de Geus}, E.~J., {de Zeeuw}, P.~T., \& {Lub}, J. 1989, \aap, 216, 44

\bibitem[{{de Zeeuw} {et~al.}(1999){de Zeeuw}, {Hoogerwerf}, {de Bruijne},
  {Brown}, \& {Blaauw}}]{deZeeuw_99_AJ}
{de Zeeuw}, P.~T., {Hoogerwerf}, R., {de Bruijne}, J.~H.~J., {Brown}, A.~G.~A.,
  \& {Blaauw}, A. 1999, \aj, 117, 354

\bibitem[{{Drew} {et~al.}(2021){Drew}, {Mongui{\'o}}, \&
  {Wright}}]{Drew_21_MNRAS}
{Drew}, J.~E., {Mongui{\'o}}, M., \& {Wright}, N.~J. 2021, \mnras, 508, 4952

\bibitem[{{Edenhofer} {et~al.}(2023){Edenhofer}, {Zucker}, {Frank}, {Saydjari},
  {Speagle}, {Finkbeiner}, \& {En{\ss}lin}}]{Edenhofer_23_arXiv}
{Edenhofer}, G., {Zucker}, C., {Frank}, P., {et~al.} 2023, arXiv e-prints,
  arXiv:2308.01295

\bibitem[{{Esplin} \& {Luhman}(2022)}]{Esplin_22_AJ}
{Esplin}, T.~L. \& {Luhman}, K.~L. 2022, \aj, 163, 64

\bibitem[{{Fern{\'a}ndez} {et~al.}(2008){Fern{\'a}ndez}, {Figueras}, \&
  {Torra}}]{Fernandez_08_AA}
{Fern{\'a}ndez}, D., {Figueras}, F., \& {Torra}, J. 2008, \aap, 480, 735

\bibitem[{{Gaia Collaboration} {et~al.}(2016){Gaia Collaboration}, {Prusti},
  {de Bruijne}, {Brown}, {Vallenari}, {Babusiaux}, \&
  {Bailer-Jones}}]{Gaia_16_AA}
{Gaia Collaboration}, {Prusti}, T., {de Bruijne}, J.~H.~J., {et~al.} 2016,
  \aap, 595, A1

\bibitem[{{Gaia Collaboration} {et~al.}(2023){Gaia Collaboration}, {Valinieri},
  {Brown}, {Prusti}, {de Bruijne}, {Arenou}, {Babusiaux}, \&
  {Biermann}}]{Gaia_23_AA}
{Gaia Collaboration}, {Valinieri}, A., {Brown}, A.~G.~A., {et~al.} 2023, \aap,
  674, A1

\bibitem[{{Galli} {et~al.}(2020){Galli}, {Bouy}, {Olivares, J.}, {Miret-Roig,
  N.}, {Sarro, L. M.}, {Barrado, D.}, {Berihuete, A.}, \& {Brandner,
  W.}}]{Galli_20_AA}
{Galli}, P. A.~B., {Bouy}, H., {Olivares, J.}, {et~al.} 2020, A\&A, 634, A98

\bibitem[{{Girichidis} {et~al.}(2018){Girichidis}, {Seifried}, {Naab},
  {Peters}, {Walch}, {W{\"u}nsch}, {Glover}, \&
  {Klessen}}]{Girichidis_18_MNRAS}
{Girichidis}, P., {Seifried}, D., {Naab}, T., {et~al.} 2018, \mnras, 480, 3511

\bibitem[{{Gratton} {et~al.}(2023){Gratton}, {Squicciarini}, {Nascimbeni},
  {Janson}, {Reffert}, {Meyer}, {Delorme}, {Mamajek}, {Bonavita}, {Desidera},
  {Mesa}, {Rigliaco}, {D'Orazi}, {Lazzoni}, {Chauvin}, \&
  {Langlois}}]{Gratton_23_arXiv}
{Gratton}, R., {Squicciarini}, V., {Nascimbeni}, V., {et~al.} 2023, arXiv
  e-prints, arXiv:2308.09962

\bibitem[{{GRAVITY Collaboration} {et~al.}(2018){GRAVITY Collaboration},
  {Abuter}, {Amorim}, {Anugu}, {Baub{\"o}ck}, {Benisty}, {Berger}, {Blind}, \&
  {Bonnet}}]{GravityColl_18_AA}
{GRAVITY Collaboration}, {Abuter}, R., {Amorim}, A., {et~al.} 2018, \aap, 615,
  L15

\bibitem[{{Gro{\ss}schedl} {et~al.}(2021){Gro{\ss}schedl}, {Alves}, {Meingast},
  \& {Herbst-Kiss}}]{Grossschedl_21_AA}
{Gro{\ss}schedl}, J.~E., {Alves}, J., {Meingast}, S., \& {Herbst-Kiss}, G.
  2021, \aap, 647, A91

\bibitem[{{Hacar} {et~al.}(2016){Hacar}, {Alves}, {Forbrich}, {Meingast},
  {Kubiak}, \& {Gro{\ss}schedl}}]{Hacar_16_AA}
{Hacar}, A., {Alves}, J., {Forbrich}, J., {et~al.} 2016, \aap, 589, A80

\bibitem[{{Haid} {et~al.}(2016){Haid}, {Walch}, {Naab}, {Seifried}, {Mackey},
  \& {Gatto}}]{Haid_16_MNRAS}
{Haid}, S., {Walch}, S., {Naab}, T., {et~al.} 2016, \mnras, 460, 2962

\bibitem[{Harris {et~al.}(2020)Harris, Millman, van~der Walt, Gommers,
  Virtanen, Cournapeau, Wieser, Taylor, Berg, Smith, Kern, Picus, Hoyer, van
  Kerkwijk, Brett, Haldane, del R{\'{i}}o, Wiebe, Peterson,
  G{\'{e}}rard-Marchant, Sheppard, Reddy, Weckesser, Abbasi, Gohlke, \&
  Oliphant}]{numpy_20}
Harris, C.~R., Millman, K.~J., van~der Walt, S.~J., {et~al.} 2020, Nature, 585,
  357

\bibitem[{{Herrington} {et~al.}(2023){Herrington}, {Dobbs}, \&
  {Bending}}]{Herrington_23_MNRAS}
{Herrington}, N.~P., {Dobbs}, C.~L., \& {Bending}, T. J.~R. 2023, \mnras, 521,
  5712

\bibitem[{Hunter(2007)}]{matplotlib_07}
Hunter, J.~D. 2007, Computing in Science \& Engineering, 9, 90

\bibitem[{{Iffrig} \& {Hennebelle}(2015)}]{Iffrig_Hennebelle_15_AA}
{Iffrig}, O. \& {Hennebelle}, P. 2015, \aap, 576, A95

\bibitem[{Inc.(2015)}]{plotly}
Inc., P.~T. 2015, Collaborative data science

\bibitem[{{Janson} {et~al.}(2021){Janson}, {Squicciarini}, {Delorme},
  {Gratton}, {Bonnefoy}, {Reffert}, {Mamajek}, {Eriksson}, {Vigan}, {Langlois},
  {Engler}, {Chauvin}, {Desidera}, {Mayer}, {Marleau}, {Bohn}, {Samland},
  {Meyer}, {d'Orazi}, {Henning}, {Quanz}, {Kenworthy}, \&
  {Carson}}]{Janson_21_AA}
{Janson}, M., {Squicciarini}, V., {Delorme}, P., {et~al.} 2021, \aap, 646, A164

\bibitem[{{Katz} {et~al.}(2023){Katz}, {Sartoretti}, {Guerrier}, {Panuzzo},
  {Seabroke}, {Th{\'e}venin}, {Cropper}, {Benson}, {Blomme}, {Haigron},
  {Marchal}, {Smith}, {Baker}, {Chemin}, {Damerdji}, {David}, {Dolding},
  {Fr{\'e}mat}, {Gosset}, {Jan{\ss}en}, {Jasniewicz}, {Lobel}, {Plum},
  {Samaras}, {Snaith}, {Soubiran}, {Vanel}, {Zwitter}, {Antoja}, {Arenou},
  {Babusiaux}, {Brouillet}, {Caffau}, {Di Matteo}, {Fabre}, {Fabricius},
  {Fragkoudi}, {Haywood}, {Huckle}, {Hottier}, {Lasne}, {Leclerc},
  {Mastrobuono-Battisti}, {Royer}, {Teyssier}, {Zorec}, {Crifo}, {Jean-Antoine
  Piccolo}, {Turon}, \& {Viala}}]{Katz_23_AA}
{Katz}, D., {Sartoretti}, P., {Guerrier}, A., {et~al.} 2023, \aap, 674, A5

\bibitem[{{Kerr} \& {Lynden-Bell}(1986)}]{Kerr_LyndenBell_86_HiA}
{Kerr}, F.~J. \& {Lynden-Bell}, D. 1986, Highlights of Astronomy, 7, 889

\bibitem[{{Kerr} {et~al.}(2021){Kerr}, {Rizzuto}, {Kraus}, \&
  {Offner}}]{Kerr_21_ApJ}
{Kerr}, R. M.~P., {Rizzuto}, A.~C., {Kraus}, A.~L., \& {Offner}, S. S.~R. 2021,
  \apj, 917, 23

\bibitem[{{Kim} \& {Ostriker}(2015)}]{Kim_Ostriker_15_ApJ}
{Kim}, C.-G. \& {Ostriker}, E.~C. 2015, \apj, 802, 99

\bibitem[{{Kirchschlager} {et~al.}(2023){Kirchschlager}, {Schmidt}, {Barlow},
  {De Looze}, \& {Sartorio}}]{Kirchschlager_23_MNRAS}
{Kirchschlager}, F., {Schmidt}, F.~D., {Barlow}, M.~J., {De Looze}, I., \&
  {Sartorio}, N.~S. 2023, \mnras, 520, 5042

\bibitem[{{Kudritzki} {et~al.}(1999){Kudritzki}, {Puls}, {Lennon}, {Venn},
  {Reetz}, {Najarro}, {McCarthy}, \& {Herrero}}]{Kudritzki_99_AA}
{Kudritzki}, R.~P., {Puls}, J., {Lennon}, D.~J., {et~al.} 1999, \aap, 350, 970

\bibitem[{{Kuhn} {et~al.}(2019){Kuhn}, {Hillenbrand}, {Sills}, {Feigelson}, \&
  {Getman}}]{Kuhn_19_ApJ}
{Kuhn}, M.~A., {Hillenbrand}, L.~A., {Sills}, A., {Feigelson}, E.~D., \&
  {Getman}, K.~V. 2019, \apj, 870, 32

\bibitem[{{Leike} {et~al.}(2020){Leike}, {Glatzle}, \&
  {En{\ss}lin}}]{Leike_20_AA}
{Leike}, R.~H., {Glatzle}, M., \& {En{\ss}lin}, T.~A. 2020, \aap, 639, A138

\bibitem[{Magnusson {et~al.}(2020)Magnusson, Vehtari, Jonasson, \&
  Andersen}]{Magnusson_20_pmlr}
Magnusson, M., Vehtari, A., Jonasson, J., \& Andersen, M. 2020, in Proceedings
  of Machine Learning Research, Vol. 108, Proceedings of the Twenty Third
  International Conference on Artificial Intelligence and Statistics, ed.
  S.~Chiappa \& R.~Calandra (PMLR), 341--351

\bibitem[{{Majewski} {et~al.}(2017){Majewski}, {Schiavon}, {Frinchaboy},
  {Allende Prieto}, {Barkhouser}, {Bizyaev}, {Blank}, {Brunner}, {Burton},
  {Carrera}, {Chojnowski}, {Cunha}, {Epstein}, {Fitzgerald}, {Garc{\'\i}a
  P{\'e}rez}, {Hearty}, {Henderson}, {Holtzman}, {Johnson}, {Lam}, {Lawler},
  {Maseman}, {M{\'e}sz{\'a}ros}, {Nelson}, {Nguyen}, {Nidever}, {Pinsonneault},
  {Shetrone}, {Smee}, {Smith}, {Stolberg}, {Skrutskie}, {Walker}, {Wilson},
  {Zasowski}, {Anders}, {Basu}, {Beland}, {Blanton}, {Bovy}, {Brownstein},
  {Carlberg}, {Chaplin}, {Chiappini}, {Eisenstein}, {Elsworth}, {Feuillet},
  {Fleming}, {Galbraith-Frew}, {Garc{\'\i}a}, {Garc{\'\i}a-Hern{\'a}ndez},
  {Gillespie}, {Girardi}, {Gunn}, {Hasselquist}, {Hayden}, {Hekker}, {Ivans},
  {Kinemuchi}, {Klaene}, {Mahadevan}, {Mathur}, {Mosser}, {Muna}, {Munn},
  {Nichol}, {O'Connell}, {Parejko}, {Robin}, {Rocha-Pinto}, {Schultheis},
  {Serenelli}, {Shane}, {Silva Aguirre}, {Sobeck}, {Thompson}, {Troup},
  {Weinberg}, \& {Zamora}}]{ApogeeOverview_17_AJ}
{Majewski}, S.~R., {Schiavon}, R.~P., {Frinchaboy}, P.~M., {et~al.} 2017, \aj,
  154, 94

\bibitem[{{Mamajek} \& {Bell}(2014)}]{Mamajek_14_MNRAS}
{Mamajek}, E.~E. \& {Bell}, C. P.~M. 2014, \mnras, 445, 2169

\bibitem[{{Mamajek} \& {Feigelson}(2001)}]{Mamajek_01_ASPC}
{Mamajek}, E.~E. \& {Feigelson}, E.~D. 2001, in Astronomical Society of the
  Pacific Conference Series, Vol. 244, Young Stars Near Earth: Progress and
  Prospects, ed. R.~{Jayawardhana} \& T.~{Greene}, 104--115

\bibitem[{McCray(1983)}]{McCray_83}
McCray, R. 1983, Highlights of Astronomy, 6, 565

\bibitem[{{Miret-Roig} {et~al.}(2020){Miret-Roig}, {Galli}, {Brandner}, {Bouy},
  {Barrado}, {Olivares}, {Antoja}, {Romero-G{\'o}mez}, {Figueras}, \&
  {Lillo-Box}}]{MiretRoig_20_AA}
{Miret-Roig}, N., {Galli}, P.~A.~B., {Brandner}, W., {et~al.} 2020, \aap, 642,
  A179

\bibitem[{{Nelson} \& {Widrow}(2022)}]{Nelson_Widrow_22_MNRAS}
{Nelson}, P. \& {Widrow}, L.~M. 2022, \mnras, 516, 5429

\bibitem[{{Neuh{\"a}user} \& {Forbrich}(2008)}]{Neuhauser_Forbich_08}
{Neuh{\"a}user}, R. \& {Forbrich}, J. 2008, in Handbook of Star Forming
  Regions, Volume II, ed. B.~{Reipurth}, Vol.~5, 735

\bibitem[{{Oort} \& {Spitzer}(1955)}]{Oort_Spitzer_55_ApJ}
{Oort}, J.~H. \& {Spitzer}, Lyman, J. 1955, \apj, 121, 6

\bibitem[{{Planck Collaboration XI}(2014)}]{Planck_14_AA}
{Planck Collaboration XI}. 2014, A\&A, 571, A11

\bibitem[{{Quintana} \& {Wright}(2021)}]{Quintana_21_MNRAS}
{Quintana}, A.~L. \& {Wright}, N.~J. 2021, \mnras, 508, 2370

\bibitem[{{Quintana} \& {Wright}(2022)}]{Quintana_22_MNRAS}
{Quintana}, A.~L. \& {Wright}, N.~J. 2022, \mnras, 515, 687

\bibitem[{{Ratzenb{\"o}ck} {et~al.}(2023b){Ratzenb{\"o}ck}, {Gro{\ss}schedl},
  {Alves}, {Miret-Roig}, {Bomze}, {Forbes}, {Goodman}, {Hacar}, {Lin},
  {Meingast}, {M{\"o}ller}, {Piecka}, {Posch}, {Rottensteiner}, {Swiggum}, \&
  {Zucker}}]{Ratzenboeck_23b_AA}
{Ratzenb{\"o}ck}, S., {Gro{\ss}schedl}, J.~E., {Alves}, J., {et~al.} 2023b,
  \aap, 678, A71

\bibitem[{{Ratzenb{\"o}ck} {et~al.}(2023a){Ratzenb{\"o}ck}, {Gro{\ss}schedl},
  {M{\"o}ller}, {Alves}, {Bomze}, \& {Meingast}}]{Ratzenboeck_23a_AA}
{Ratzenb{\"o}ck}, S., {Gro{\ss}schedl}, J.~E., {M{\"o}ller}, T., {et~al.}
  2023a, \aap, 677, A59

\bibitem[{{Rizzuto} {et~al.}(2011){Rizzuto}, {Ireland}, \&
  {Robertson}}]{Rizzuto_11_MNRAS}
{Rizzuto}, A.~C., {Ireland}, M.~J., \& {Robertson}, J.~G. 2011, \mnras, 416,
  3108

\bibitem[{{Rogers} \& {Pittard}(2013)}]{Rogers_Pittard_13_MNRAS}
{Rogers}, H. \& {Pittard}, J.~M. 2013, \mnras, 431, 1337

\bibitem[{{Rogers} \& {Pittard}(2014)}]{Rogers_Pittard_14_MNRAS}
{Rogers}, H. \& {Pittard}, J.~M. 2014, \mnras, 441, 964

\bibitem[{{R{\"o}ser} {et~al.}(2018){R{\"o}ser}, {Schilbach}, {Goldman},
  {Henning}, {Moor}, \& {Derekas}}]{Roeser_18_AA}
{R{\"o}ser}, S., {Schilbach}, E., {Goldman}, B., {et~al.} 2018, \aap, 614, A81

\bibitem[{Salvatier {et~al.}(2016)Salvatier, Wiecki, \& Fonnesbeck}]{PyMC_16}
Salvatier, J., Wiecki, T.~V., \& Fonnesbeck, C. 2016, {PeerJ} Computer Science,
  2, e55

\bibitem[{{Sandell} {et~al.}(2021){Sandell}, {Reipurth}, {Vacca}, \&
  {Bajaj}}]{Sandell_21_ApJ}
{Sandell}, G., {Reipurth}, B., {Vacca}, W.~D., \& {Bajaj}, N.~S. 2021, \apj,
  920, 7

\bibitem[{{Sch{\"o}nrich} {et~al.}(2010){Sch{\"o}nrich}, {Binney}, \&
  {Dehnen}}]{Schoenrich_10_MNRAS}
{Sch{\"o}nrich}, R., {Binney}, J., \& {Dehnen}, W. 2010, \mnras, 403, 1829

\bibitem[{{Taylor} \& {Storey}(1984)}]{Taylor_Storey_84_MNRAS}
{Taylor}, K.~N.~R. \& {Storey}, J.~W.~V. 1984, \mnras, 209, 5P

\bibitem[{{Tobin} {et~al.}(2009){Tobin}, {Hartmann}, {Furesz}, {Mateo}, \&
  {Megeath}}]{Tobin_09_ApJ}
{Tobin}, J.~J., {Hartmann}, L., {Furesz}, G., {Mateo}, M., \& {Megeath}, S.~T.
  2009, \apj, 697, 1103

\bibitem[{{Torres} {et~al.}(2006){Torres}, {Quast}, {da Silva}, {de La Reza},
  {Melo}, \& {Sterzik}}]{Torres_06_AA}
{Torres}, C.~A.~O., {Quast}, G.~R., {da Silva}, L., {et~al.} 2006, \aap, 460,
  695

\bibitem[{Virtanen {et~al.}(2020)Virtanen, Gommers, Oliphant, Haberland, Reddy,
  Cournapeau, Burovski, Peterson, Weckesser, Bright, {van der Walt}, Brett,
  Wilson, Millman, Mayorov, Nelson, Jones, Kern, Larson, Carey, Polat, Feng,
  Moore, {VanderPlas}, Laxalde, Perktold, Cimrman, Henriksen, Quintero, Harris,
  Archibald, Ribeiro, Pedregosa, {van Mulbregt}, \& {SciPy 1.0
  Contributors}}]{Scipy_20_NMeth}
Virtanen, P., Gommers, R., Oliphant, T.~E., {et~al.} 2020, Nature Methods, 17,
  261

\bibitem[{{Walch} {et~al.}(2015){Walch}, {Girichidis}, {Naab}, {Gatto},
  {Glover}, {W{\"u}nsch}, {Klessen}, {Clark}, {Peters}, {Derigs}, \&
  {Baczynski}}]{Walch_15_MNRAS}
{Walch}, S., {Girichidis}, P., {Naab}, T., {et~al.} 2015, \mnras, 454, 238

\bibitem[{{Walch} \& {Naab}(2015)}]{Walch_Naab_15_MNRAS}
{Walch}, S. \& {Naab}, T. 2015, \mnras, 451, 2757

\bibitem[{Watanabe(2013)}]{Watanabe_13_JMLR}
Watanabe, S. 2013, J. Mach. Learn. Res., 14, 867–897

\bibitem[{{Zucker} {et~al.}(2022){Zucker}, {Goodman}, {Alves}, {Bialy},
  {Foley}, {Speagle}, {Gro{\^I}{\texttwosuperior}schedl}, {Finkbeiner},
  {Burkert}, {Khimey}, \& {Swiggum}}]{Zucker_22_Natur}
{Zucker}, C., {Goodman}, A.~A., {Alves}, J., {et~al.} 2022, \nat, 601, 334

\bibitem[{{Zucker} {et~al.}(2020){Zucker}, {Speagle}, {Schlafly}, {Green},
  {Finkbeiner}, {Goodman}, \& {Alves}}]{Zucker_20_AA}
{Zucker}, C., {Speagle}, J.~S., {Schlafly}, E.~F., {et~al.} 2020, \aap, 633,
  A51

\end{thebibliography}


\begin{appendix}

\section{Auxiliary Tables and Figures}\label{app:clusterprop}

    In this appendix section, we provide additional material to support the findings we presented in this letter.

    Table \ref{tab:properties} lists spatial and kinematic properties as well as ages of the seven clusters we discuss.

    \begin{table*}[t]
    \centering
    \caption{Summary of the average positional and kinematic properties of the clusters that were initially selected to be part of the CrA chain.}
    \label{tab:properties}
        \resizebox{1\textwidth}{!}{%
        \begin{tabular}{c|c|ccccccc}
        \toprule \toprule
                                                                                             &                      & CrA Main      & CrA North     & Sco Sting     & Sco Body      & V1062 Sco & $\eta$ Lupus & $\phi$ Lupus
        \\
        \midrule\relax
        \multirow{4}{*}{\begin{tabular}[c]{@{}l@{}}number\\of stars\end{tabular}}            & total                & 96            & 351           & 132           & 373           & 1029          & 769           & 1114         \\
                                                                                             & RVs                  & 56 (23) (33)  & 134 (126) (8) & 42 (40) (2)   & 143 (139) (4) & 309 (309) (0) & 360 (357) (3) & 455 (455) (0)\\
                                                                                             & eRV$<$3              & 24 (0) (24)   & 32 (26) (6)   & 7 (6) (1)     & 25 (25) (0)   & 58 (58) (0)   & 69 (68) (1)   & 86 (86) (0)  \\
                                                                                             & eRV$<$1              & 22 (0) (22)   & 10 (6) (4)    & 5 (4) (1)     & 10 (10) (0)   & 17 (17) (0)   & 30 (29) (1)   & 38 (38) (0)  \\
        \midrule\relax                                                                           
        \multirow{6}{*}{\begin{tabular}[c]{@{}l@{}}hel Cart. \\coord. [pc]\end{tabular}}     & X                    & 148           & 145           & 132           & 139           & 167           & 124           & 112          \\
                                                                                             & $\sigma_\mathrm{X}$  & 3             & 5             & 11            & 10            & 7             & 9             & 13           \\
                                                                                             & Y                    & -1            & -3            & -21           & -26           & -54           & -45           & -54          \\
                                                                                             & $\sigma_\mathrm{Y}$  & 2             & 4             & 8             & 7             & 9             & 13            & 9            \\
                                                                                             & Z                    & -47           & -35           & -7            & 19            & 15            & 25            & 40           \\
                                                                                             & $\sigma_\mathrm{Z}$  & 2             & 7             & 9             & 12            & 5             & 9             & 9            \\
        \midrule\relax                                                                                             
        [pc]                                                                                 & d$^{\ast}$           & 155           & 149           & 134           & 141           & 177           & 136           & 131          \\
        \midrule\relax                                                                                            
        \multirow{9}{*}{\begin{tabular}[c]{@{}l@{}}hel Cart. \\vel. [km/s]\end{tabular}}     & U                    & -4.5          & -5.5          & -5.9          & -3.1          & -3.9          & -4.6          & -5.0         \\
                                                                                             & $\sigma_\mathrm{U}$  & 1.2           & 1.4           & 2.3           & 1.6           & 1.5           & 1.5           & 3.8          \\
                                                                                             & eU$_\mathrm{avg}$    & 0.3           & 0.6           & 0.5           & 0.7           & 0.7           & 0.5           & 0.5          \\
                                                                                             & V                    & -17.8         & -17.4         & -19.0         & -16.8         & -19.6         & -19.9         & -19.8        \\
                                                                                             & $\sigma_\mathrm{V}$  & 0.8           & 0.6           & 0.5           & 0.6           & 0.4           & 1.2           & 1.7          \\
                                                                                             & eV$_\mathrm{avg}$    & 0.3           & 0.1           & 0.1           & 0.1           & 0.2           & 0.2           & 0.3          \\
                                                                                             & W                    & -9.7          & -7.8          & -5.5          & -6.9          & -4.2          & -5.2          & -4.9         \\
                                                                                             & $\sigma_\mathrm{W}$  & 0.7           & 1.0           & 0.2           & 0.7           & 0.3           & 0.6           & 2.1          \\
                                                                                             & eW$_\mathrm{avg}$    & 0.2           & 0.2           & 0.0           & 0.1           & 0.1           & 0.1           & 0.2          \\
        \midrule\relax                                                                                             
        [Myr]                                                                                & age$^{\ast\ast}$     & 8.5           & 11.9          & 14.5          & 14.7          & 15.0          & 15.3          & 16.9         \\
        \bottomrule                                                                                             
        \end{tabular}%
    }
    \tablefoot{
        Rows 1--3 give the number of stars per cluster in total that have radial velocities, and how many stars remain after rejecting binary candidates and applying our error cuts of $<$~\SI{3}{\km\per\s} and $<$~\SI{1}{\km\per\s}, respectively. In parentheses, we list the number of stars with radial velocities first from \textit{Gaia} and second from APOGEE-2. The following statistics of the data are based on the heliocentric Cartesian coordinate system. Rows 4--9 list the mean positions and standard deviation per cluster, excluding binary candidates, low-stability stars, and sigma outliers, but including stars without radial velocity measurements. The cluster distance is taken directly from \citetalias{Ratzenboeck_23a_AA}. The mean standard deviation and average uncertainty of the velocities (rows 12--20) are taken from our sample of stars with radial velocity errors $<$~\SI{1}{\km\per\s}. Ages are listed as derived by \citetalias{Ratzenboeck_23b_AA}.
    }
    \end{table*}

\begin{figure}[h]
        \centering
        \includegraphics[width=\columnwidth]{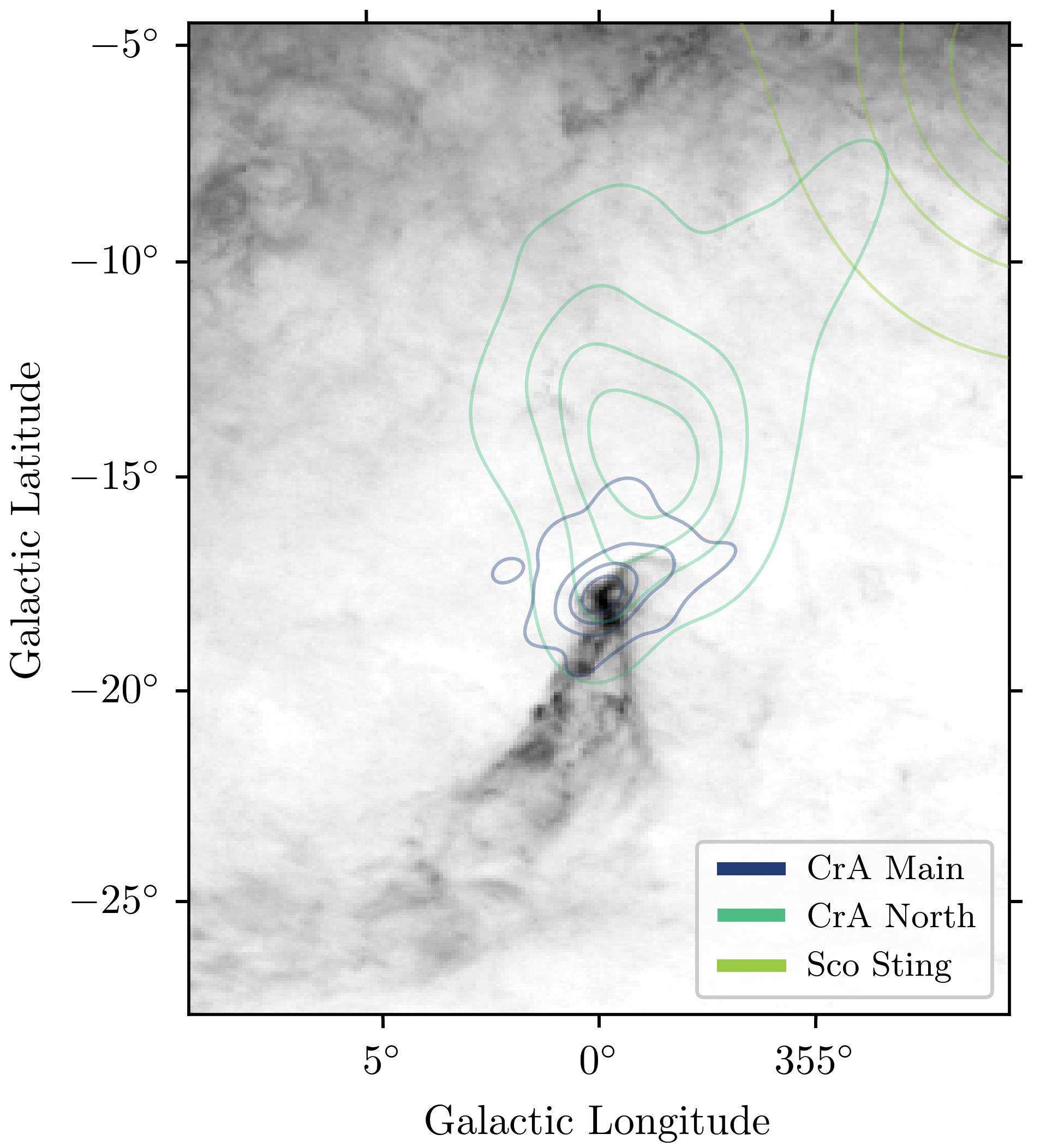}
        \caption{Cutout of the Planck E(B-V) dust map \citep{Planck_14_AA} showing the wind-blown appearance of the CrA molecular cloud. We overplot stellar densities of the clusters CrA~Main, North, and Sco~Sting with contour lines for reference.}
        \label{fig:planck}
    \end{figure}

    Figure~\ref{fig:planck} shows the wind-blown shape of the CrA molecular cloud and the relative position of the CrA clusters in Galactic coordinates. The top of the cloud is denser, possibly influenced by a feedback force originating from the Galactic north and compressing the molecular cloud. The clusters are indicated with stellar density contours (\num{5}\%, \num{25}\%, \num{50}\%, and \num{75}\%) on top.
    
\section{Data selection}\label{app:data}

    We initially included all the stars that were identified as part of the CrA chain by \citetalias{Ratzenboeck_23a_AA}, but removed outliers with the \verb|stability| criterion from \citetalias{Ratzenboeck_23a_AA} as described in Sect.~\ref{sec:data}. We then combined radial velocity data from \textit{Gaia} DR3 \citep{Gaia_23_AA, Katz_23_AA} and APOGEE-2 from SDSS DR17 \citep{Apogee2_22_ApJ}.

    For the APOGEE-2 radial velocity error, we added the nominal uncertainty (\texttt{VERR}) in quadrature to the scatter for stars with more than one visit (\texttt{VSCATTER}). We rejected APOGEE-2 stars with \verb|VSCATTER| values higher than \SI{3}{\kilo\meter\per\second} as well as stars whose radial velocity measurement difference between \textit{Gaia} and APOGEE-2 was larger than \SI{3}{\km\per\s}. When the \textit{Gaia} DR3 radial velocity errors of the stars were larger than the measurement difference between the \textit{Gaia} and APOGEE-2 radial velocities, we kept them in the sample. Both criteria reject binary star candidates or poor measurements that would impede a precise kinematic analysis. We used the measurement with the smallest error for stars with both \textit{Gaia} and APOGEE-2 radial velocity measurements.

    As a next step, we excluded stars based on their radial velocity errors. We restricted our sample for the kinematic analysis to stars with radial velocity errors $<$~\SI{1}{\km\per\s} and for the traceback analysis to stars with radial velocity errors $<$~\SI{3}{\km\per\s}. The number of stars with radial velocities per cluster before and after the error cuts is listed in Table~\ref{tab:properties}.

    We transformed the \textit{Gaia} DR3 astrometry and our restricted selection of radial velocities to heliocentric Galactic Cartesian coordinates and velocities (\textit{X,~Y,~Z,~U,~V,~W}) using \texttt{astropy SkyCoord} \citep{Astropy_22_ApJ}. We employed a Monte Carlo approach to propagate the uncertainties from observed coordinates into heliocentric Galactic Cartesian coordinates, sampling from joint normal distributions centered on the observed measurements and corresponding astrometric covariances. We combined the 5D astrometric covariance information provided by \textit{Gaia} with the radial velocity uncertainty information, assuming no correlation between the two subspaces. The final Cartesian coordinates, velocities, and corresponding uncertainties are the means and standard deviations of 1\,000 samples per star.

    After the conversion into heliocentric Galactic Cartesian coordinates, we further rejected 12, 9, and 10 stars for the more massive clusters V1062~Sco, $\eta$~Lupus, and $\phi$~Lupus, respectively. This is based on 2-sigma clipping in all velocity-velocity spaces (\textit{U}~vs.~\textit{V}, \textit{U}~vs.~\textit{W}, and \textit{V}~vs.~\textit{W}) for the sample of stars with radial velocity errors $<$~\SI{3}{\km\per\s}. We note that the line of sight to CrA almost coincides with the Galactic Cartesian \textit{X}-axis, and therefore, the \textit{U}-velocity is dominated by the radial velocity uncertainty. These uncertainties are larger than the \textit{V}- and \textit{W}-velocity uncertainties because the \textit{Gaia} proper motion measurements with a lower uncertainty dominate the latter. Additionally, CrA Main exhibits larger than average uncertainties in its \textit{X}-coordinate, likely due to the higher extinction from the CrA molecular cloud that compromises precise parallax measurements.

    The average position and velocity uncertainties per cluster are given in Table \ref{tab:properties}.

\section{LSR-corrected radial velocity and momentum estimation}\label{app:LoSMom}

    We estimated the current momentum of the CrA molecular cloud to infer the necessary number of SN explosions needed to explain the cloud momentum. The estimation was carried out using a Monte Carlo-type sampling of the parameters.

    \begin{figure}[h]
        \centering
        \includegraphics[width=\columnwidth]{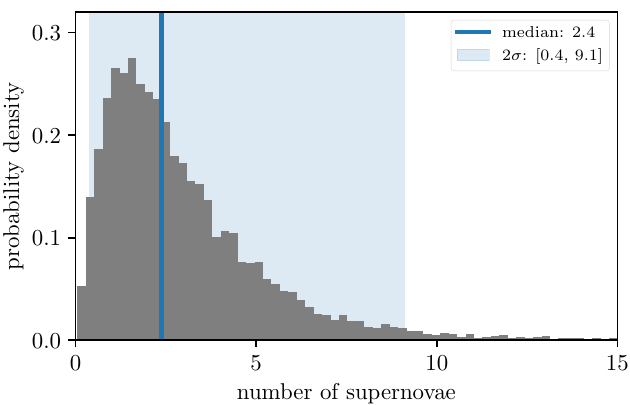}
        \caption{Probability distribution of the number of SN explosions needed to explain the current momentum of the CrA molecular cloud. The median number of SNe necessary is about two, with \num{95}\% of values lying between \num{0.4} and \num{9.1}.}
        \label{fig:supernovae}
    \end{figure}

    \subsection{LSR-corrected radial velocity}
    
    We find that CrA~Main matches the velocity of the CO gas in the molecular cloud. For this comparison, we converted the mean heliocentric radial velocity of CrA~Main (\SI{-1.3\pm1.3}{\km\per\s}) into the radial velocity relative to the local standard of rest (LSR).
    For the conversion, we employed a Monte Carlo-type sampling with marginal normal distributions based on the observables and their uncertainties. 
    Because the measurement of the solar motion from \cite{Kerr_LyndenBell_86_HiA} defined the LSR when \cite{Dame_01_ApJ} published the composite CO survey of the CrA molecular cloud, we used this definition of the LSR for the radial velocity comparison. As a result, we determined a radial velocity relative to the LSR of \SI{5.9\pm0.3}{\km\per\s}. Within the uncertainties, this value is consistent with the radial velocity measurement corrected for solar motion of the CO gas within the CrA molecular cloud, as inferred by \cite{Dame_01_ApJ} at \SI{5.7\pm1.4}{\km\per\s}.
    The consistency in the motion between CrA Main and the CO gas within the molecular cloud suggests that the young stars within the CrA star formation complex retain the motion history of the cloud for several \si{\mega\year} \citep[see also][]{Grossschedl_21_AA}. This supports our use of the motion of the CrA Main cluster as an approximation for the current motion of the CrA molecular cloud.

    \subsection{Momentum estimation}
    First, we calculated the current momentum of the CrA molecular cloud, before we estimated how many SN explosions or how much stellar wind we need to achieve the estimated cloud momentum.

    As a mass estimate of the CrA molecular cloud, we used the cloud masses inferred by \cite{Alves_14_AA}\footnote{To infer the mass from 2MASS K-band extinction maps the authors used $M = d^2 \mu \beta_{K} \int_{\Omega} A_K (\Vec{\Theta}) d^2 \Theta$}, \SI{7060}{\solarmass} as total mass, \SI{4200}{\solarmass} as mass with $A_K > 0.1$~mag, and \SI{950}{\solarmass} as mass with $A_K > 0.2$~mag.
    The extinction thresholds correspond to different gas densities. $A_K \gtrsim 0.16$~mag is the critical density threshold for forming molecular CO gas \citep[e.g.,][]{Alves_99_ApJ}.
    We corrected these estimates for the recently improved cloud distance. A$_K$ represents the extinction in the K band observed by the authors,

    \begin{equation}
        f_\mathrm{correction} = \left(d_\mathrm{Z20} / d_\mathrm{A14}\right)^2,
    \end{equation}
    
    with the old distance measurement d$_\mathrm{A14}~=$~\SI{130}{\parsec} \citep{Casey_98_AJ, Alves_14_AA} and the updated distance d$_\mathrm{Z20}~=$~\SI{150}{\parsec} \citep{Zucker_20_AA}. The corrected mass was computed by multiplying the old mass by the correction factor.

    We obtained a corrected mass of \SI{9500}{\solarmass} for the total cloud mass (\SI{5700}{\solarmass} and \SI{1300}{\solarmass} for the denser gas). We generated 10~000 values from a gamma distribution with parameters that were determined based on the mean (\SI{5700}{\solarmass}) and standard deviation (\SI{3400}{\solarmass}) of the corrected mass estimates. Applying the gamma function ensured that all sampled values are greater than zero.

    The velocity of CrA~Main was used as an approximation of the current absolute velocity of the cloud.
    We used three different reference frames as rest velocity. Each reference cluster was selected as a potential source of feedback, thus showing the difference in the motion of the CrA molecular cloud to that of the potential feedback source. First, we used the motion of $\phi$~Lupus ((UVW)$_\mathrm{LSR}$~=~(-5.7, -19.5, -5.1)) as rest frame because it is the oldest cluster in the CrA chain and a potential source of feedback. As a second rest frame, we used the motion of V1062~Sco ((UVW)$_\mathrm{LSR}$~=~(-5.5, -19.7, -5.0)) because this cluster is a possible feedback source as well. As a third rest frame, we used the average motion of all stars in Sco-Cen for comparison, but excluded clusters younger than \SI{12}{\mega\year} ((UVW)$_\mathrm{LSR}$~=~(-6.0, -19.5, -5.3)). The motion of the younger clusters was likely affected by feedback as well. This resulted in relative velocities for the CrA molecular cloud of \SI{5.1}{\km\per\s}, \SI{5.2}{\km\per\s}, and \SI{5.0}{\km\per\s}, respectively. We sampled 10~000 values from a normal distribution with a mean (\SI{5.1}{\km\per\s}) and standard deviation (\SI{0.1}{\km\per\s}) of the relative velocities.
    
    We then computed the momentum as $P = M \cdot v$ and obtained a distribution of the cloud momentum with a median of \SI{25000}{\solarmass\km\per\s}.

    Next, we estimated how many SNe are needed to explain the current momentum of the cloud while considering that only a fraction of the cloud surface is exposed to the SN shell.
    Based on the tracebacks we computed for the CrA chain clusters, \SIrange{10}{15}{\mega\year} ago, CrA~North was at a distance of \SI{30}{\parsec} from the more massive clusters at the center of Sco-Cen. We used a normal distribution centered on this distance, with a standard deviation of \SI{5}{\parsec}, as the radius of the SN sphere at the time it hit the primordial CrA cloud and derived a surface area of the SN shell ($A_{sp} = 4\pi \cdot r^2$). We find a current cloud radius of roughly \SI{5}{\parsec}. Assuming the CrA molecular cloud was larger in the past, we approximated the cloud radius with a uniform distribution between \SIrange{10}{12}{\parsec}, with which we calculated a circular area ($A_c = \pi \cdot r^2$) that was exposed to the SN shell. This area accounts for a fraction of the total surface area of the sphere. The distribution peaks at about \num{0.03}, and consequently, we estimated that this fraction of the total SN momentum output was injected into the cloud.
    We used a range of $P_\mathrm{SN, total}$ = \SIrange{2e5}{4e5}{\solarmass\km\per\s} as the total SN momentum output, which we adopted from \cite[e.g.,][]{Iffrig_Hennebelle_15_AA, Kim_Ostriker_15_ApJ, Walch_Naab_15_MNRAS, Haid_16_MNRAS}.
    These values are primarily contingent on the density of the ambient medium and the location of the explosion relative to the molecular cloud. A lower-density medium and an explosion occurring within the molecular cloud result in higher momentum transfer.
    Again, we sampled 10,000 values from a uniform distribution between the given range of \SIrange{2e5}{4e5}{\solarmass\km\per\s} as we wished to refrain from assuming a predominant density or the location of the explosion relative to the molecular cloud.
    Dividing the current momentum of the cloud by the fraction of the total SN momentum output that is transferred to the cloud gives the number of SNe that are needed to account for this momentum.

    The median of the needed number of SNe is at about two SNe to explain the current momentum of the CrA molecular cloud, with a \num{95}\% probability range spanning from less than one to about nine SNe.
    
    In this statistical approach, we covered a range of parameters, but the resulting number of SNe can vary further based on some additional factors. Our assumption of \num{100}\% efficiency in transferring momentum from the SN to the cloud motion is not a realistic representation. Furthermore, before SN explosions occur, both stellar winds and photoionizing radiation erode potential surrounding molecular material away, creating channels through which the hot gas can escape. This process redirects momentum away from the surrounding clouds \citep[e.g.,][]{Rogers_Pittard_13_MNRAS, Rogers_Pittard_14_MNRAS}.

    \subsection{Stellar winds from B-type stars}\label{ssub:btype}
    Because the Sco-Cen OB association includes several B-type stars \citep[e.g.,][]{deGeus_89_AA, deZeeuw_99_AJ, Gratton_23_arXiv}, we also explored the possibility of stellar winds as a constant source for momentum injection into their environment. We averaged the observed data, their mass-loss rate, and terminal velocities of 14 B-type supergiants published by \cite{Kudritzki_99_AA} that are unrelated to the Sco-Cen region as an upper limit in our approximation. Using $P = M \cdot v$, we obtained a total momentum output of $\sim$~\SI{4600}{\solarmass\km\per\s} over a time period of \SI{10}{\mega\year}.
    We considered that the progenitor of the CrA molecular cloud is exposed to a fraction of the sphere of the stellar wind. Again, we placed a star at a distance of \SI{30}{\parsec} that exposes the cloud to about \num{3}\% of the total momentum of the stellar wind. As a result, we find that between about 50 and 350 B-type supergiants with constant winds over the past \SI{10}{\mega\year} are needed to inject the momentum the CrA molecular cloud has today. Several B-type stars lie at varying distances to the CrA molecular cloud in Sco-Cen \citep{Gratton_23_arXiv}. However, even when we placed massive stars at a distance of \SI{10}{\parsec}, the required number ranges between 6 and 40, depending on the cloud momentum. Although stellar winds likely contribute to the injected momentum, we conclude that they alone cannot cause the motion of the CrA cloud.

\section{Bayesian linear regression of the position-velocity relations and Bayesian model comparison}\label{app:model_comparison}

    We employed a Bayesian linear regression routine to the cluster data to quantify the expansion between the clusters. Using the \texttt{PyMC} \textit{Python} library \citep{PyMC_16}, we fit a linear model with 1~000 draws using relatively uninformative priors, that is, zero-centered Gaussian priors with a standard deviation of 20 on all regression parameters. Because the problem has relatively few dimensions, this does not affect the outcome of the inference. The fitting parameters are listed in Tab.~\ref{tab:fit_param}, and the fits using the median of all sampled parameters are plotted in Fig.~\ref{fig:gradient}.  
    
    \begin{table}[!h]
    \centering
    \caption{Fit parameters from a Bayesian linear regression to all clusters in the upper two panels of Fig.~\ref{fig:gradient}.}
    \label{tab:fit_param}
    \resizebox{0.65\columnwidth}{!}{%
    \begin{tabular}{cccc}
        \toprule \toprule
                                 & slope              & intercept           \\
                                 & km/s/pc            & km/s                \\
        \midrule
        U vs. X                  & 0.031 $\pm$ 0.005  & -8.927 $\pm$ 0.610  \\
        V vs. Y                  & 0.040 $\pm$ 0.002  & -17.721 $\pm$ 0.095 \\
        \bottomrule
        \end{tabular}%
    }
    \tablefoot{The Bayesian linear regression returns a posterior probability distribution for each fitting parameter; we call them ``slope'' and ``intercept'' for the linear fit. The parameter value and its uncertainty are computed as the median and median absolute deviation of the posterior probability distribution, respectively.}
    \end{table}

    \begin{figure}[t]
        \centering
        \includegraphics[width=\columnwidth]{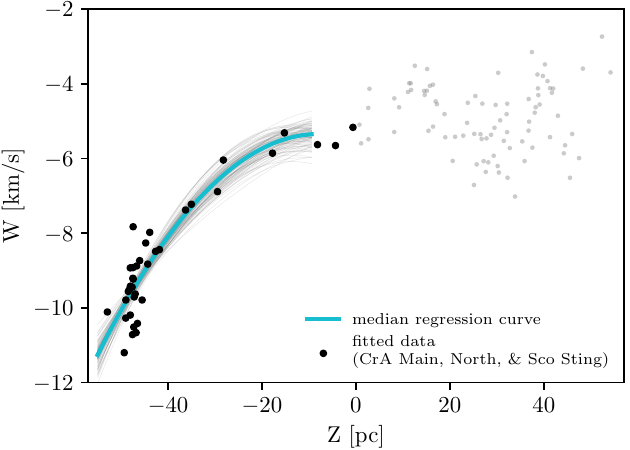}
        \caption{Bayesian quadratic fit to the stars of the clusters CrA~Main, North, and Sco~Sting, shown as black dots. The regression curve with the median of all sampled parameters is plotted as a solid cyan line, a \num{100} random fitted curves are shown as light gray lines. The Bayesian model comparison significantly prefers the quadratic over a linear fit.}
        \label{fig:quadratic}
    \end{figure}
    
    In addition to the Bayesian linear regression that we applied to all CrA chain data in all dimensions, we also fit a quadratic curve in the \textit{W}~versus~\textit{Z} space to a subsection of CrA chain clusters, namely CrA~Main, North, and Sco~Sting. We show this in Fig.~\ref{fig:quadratic} and visually find this to be a good fit. The fit function is $f(x) = -0.003x^2 - 0.044x - 5.507$.
    
    To support this conclusion quantitatively, we performed a model comparison and used a Bayesian regression code using the \texttt{PyMC} library \citep{PyMC_16}. First, we fit a linear and quadratic model to the data using the same priors as described previously. Using samples drawn from the posterior and the computed log point-wise posterior predictive density, we can subsequently apply the leave-one-out cross-validation \citep[LOO-CV,][]{Magnusson_20_pmlr} and the widely-applicable information criterion \citep[WAIC,][]{Watanabe_13_JMLR} for the model selection. This approach allows us to establish a statistical preference for either of the fitting functions applied to the data.
    Using the LOO-CV, the model is evaluated based on its expected predictive accuracy on data points that are not used in fitting the model.
    The WAIC takes both the model goodness of fit to the data and the model complexity into consideration, providing a more nuanced view of model performance.
    In the vertical dimension and for the subsection of three clusters in the CrA chain CrA~Main, North, and Sco~Sting, we find that the model comparison significantly favors the quadratic fit. This indicates that a constant feedback force affects the motion of their primordial cloud. Nevertheless, the fit does not clarify which of the older and more massive clusters, that is, V1062~Sco, or $\eta$~Lupus and $\phi$~Lupus, serves as the primary source of stellar feedback.

    \subsection{Comparing the linear position-velocity gradients to the expected Milky Way dynamics}
    
        We used the Oort constants to estimate the anticipated position-velocity gradient in the solar neighborhood, assuming that only Galactic forces affect the clusters, and compared them to the observed linear expansion gradients. Based on previous publications, \cite{Nelson_Widrow_22_MNRAS} specified $\delta\text{v}_x/\delta\text{x} \equiv$ K + C and $\delta\text{v}_y/\delta\text{y} \equiv$ K - C as the definition of the Oort constants. They computed the stellar velocity field in the vicinity of the Sun and derived values for the Oort constants from this by differentiating the velocity field. For a selected volume around the position of the Sun with a radius of \SI{500}{\parsec}, they found K and C to be \num{-2.3 \pm 0.5} and \SI{-2.9 \pm 0.8}{\km\per\s\per\kilo\parsec}, respectively. We compared the CrA chain \textit{U}~versus~\textit{X} and \textit{V}~versus~\textit{Y} gradients (Table~\ref{tab:fit_param}, $\sim$~\num{41 \pm 17} and \SI{38 \pm 4}{\km\per\s\per\kilo\parsec}, respectively) to the Galactic position-velocity gradients (\num{-5.2 \pm 0.9} and \SI{0.6 \pm 0.9}{\km\per\s\per\kilo\parsec}, respectively) and find that the former are larger by an order of magnitude. We conclude that the CrA chain expands faster than it would if it were induced by Galactic rotation alone.
        
 \section{Traceback analysis}\label{app:traceback}        

    \begin{table}[t]
    \centering
    \caption{Median of the minimum distances (Col.~3) at the respective time (Col.~4) between CrA~Main, North, Sco~Sting and the clusters V1062~Sco and $\eta$ and $\phi$~Lupus.}
    \label{tab:distances}
    \resizebox{0.8\columnwidth}{!}{%
        \begin{tabular}{llcc}
        \toprule \toprule
        distance to                     &               & d$_\mathrm{min}$  [pc] & t(d$_\mathrm{min}$)  [Myr] \\
        \midrule
        \multirow{3}{*}{V1062 Sco}      & CrA Main      & 37.5                   & -10.2                      \\
                                        & CrA North     & 28.1                   & -12.8                      \\
                                        & Sco Sting     & 31.0                   & -16.0                      \\
        \midrule
        \multirow{3}{*}{$\eta$ Lupus}   & CrA Main      & 30.7                   & -11.8                      \\
                                        & CrA North     & 28.0                   & -13.8                      \\
                                        & Sco Sting     & 26.6                   & -15.8                      \\
        \midrule
        \multirow{4}{*}{$\phi$ Lupus}   & CrA Main      & 37.1                   & -13.2                      \\
                                        & CrA North     & 30.2                   & -15.5                      \\
                                        & Sco Sting     & 30.8                   & -18.4                      \\

        \bottomrule
        \end{tabular}
    }
    \end{table}

    For the orbital tracebacks, we used the Galactic potential by \citet{galpy_15} as implemented in \texttt{galpy} (\texttt{MWPotential2014}). All tracebacks were calculated using the \texttt{Astropy 4.0} default values, including the standard solar motion relative to the LSR from \cite{Schoenrich_10_MNRAS}, (\textit{U}, \textit{V}, \textit{W})$_\mathrm{LSR}$~=~(-11.1, 12.24, 7.25) \si{\km\per\s}, and the galactocentric solar position of ($X_G$, $Y_G$, $Z_G$)~=~(8122.0, 0.0, 20.8) \si{\parsec} \citep{GravityColl_18_AA, Bennett_Bovy_19_MNRAS}. Here, we used all stars with radial velocity errors $<$~\SI{3}{\km\per\s} and bootstrapped over the cluster members with replacement to obtain \num{1000} orbits for each cluster. Then, we calculated the Euclidean distance between the clusters for all sampled orbits at time steps of \SI{0.1}{\mega\year} between \SIrange{-20}{0}{\mega\year} and calculated the median and median absolute deviation as the confidence intervals.
    Figure~\ref{fig:traceback} shows the median distances between the older and more massive clusters, V1062~Sco, $\eta$~Lupus, and $\phi$~Lupus, to each of the younger clusters, CrA~Main, CrA~North, and Sco~Sting, over time. The minimum distances and the times of minimum distance are given in Table~\ref{tab:distances}.

    It is important to note that an accelerated motion or push toward any direction would lead to an incorrect orbit computation of a cloud. When a vertical push downward from feedback is accounted for, the orbit inclination would decrease, leading to a shallower oscillation. Additionally, assuming the current motion as the constant cluster motion will lead to a position that overshoots its actual initial position because the traceback does not consider acceleration. Therefore, the given tracebacks and minimum distances between clusters should be only seen as a rough approximation, and the true uncertainties are likely larger than the given confidence intervals. In future work, we will model a scenario that includes acceleration due to massive stellar feedback to understand its effect on the cluster orbits on timescales of several \si{\mega\year}.

\end{appendix}

\end{document}